\documentclass[prb,showpacs,amsmath,amssymb,twocolumn]{revtex4}

\usepackage[pdftex]{graphicx}
\usepackage[pdftex]{epsfig}
\usepackage{epstopdf}
\usepackage{dcolumn}
\usepackage{bm}

\graphicspath{{converted_graphics/}}
\begin{document}

\title{Theory of valley-orbit coupling in a Si/SiGe quantum dot}

\author{Mark Friesen}
\email{Electronic mail:  friesen@cae.wisc.edu}
\author{S. N. Coppersmith}
\affiliation{Department of Physics, University of
Wisconsin-Madison, Wisconsin 53706, USA} 

\begin{abstract}
Electron states are studied for quantum dots in a strained Si quantum
well, taking into account both valley and orbital physics.  Realistic geometries are considered, including circular and elliptical dot shapes, parallel and perpendicular magnetic fields, and (most importantly for valley coupling) the small local tilt of the quantum well interface away from the crystallographic axes.  In absence of a tilt, valley splitting occurs only between pairs of states with the same orbital quantum numbers.  However, tilting is ubiquitous in conventional silicon heterostructures, leading to valley-orbit coupling.  In this context, ``valley splitting" is no longer a well defined concept, and the quantity of merit for qubit applications becomes the ground state gap.  For typical dots used as qubits, a rich energy spectrum emerges, as a function of magnetic field, tilt angle, and orbital quantum number.  Numerical and analytical solutions are obtained for the ground state gap and for the mixing fraction between the ground and excited states.  This mixing can lead to valley scattering, decoherence, and leakage for Si spin qubits.
\end{abstract}


\maketitle

\section{Introduction}
In the past decade, much progress has been made in the development of quantum dot devices for quantum information processing, spurred by the spin qubit proposal of Loss and DiVincenzo.\cite{loss98}  In GaAs devices, recent achievements have included single electron occupation of a quantum dot,\cite{ciorga00,fujisawa02} pulsed gating and single shot readout,\cite{elzerman04} and qubit rotations about an arbitrary axis.\cite{petta05}
However, it was noted early on that materials properties in silicon are potentially well suited for quantum computing.\cite{kane98}  Predominant among these are weak spin-orbit coupling and the ready availability of a nuclear spin-0 isotope.  

Several proposals were put forth for silicon-based quantum dot spin qubits.\cite{vrijen00,friesen03,tucker04,hill05}  Many of the technological achievements made in GaAs have now been realized in Si.  Recent experimental advances include controlled tunnel barriers in modulation doped,\cite{sakr05,slinker05,schaeffler06,angus07} degenerately doped,\cite{ruess08} and undoped enhancement-mode devices,\cite{eric1,eric2} coupling between a quantum dot and a nearby charge sensor,\cite{sakr05,buehler06,simmons07} controlled tunnel coupling between two sides of a double dot,\cite{chan06,simmons09} single electron occupation of a quantum dot,\cite{simmons07} and observations of coherent quantum phenomena involving spins.\cite{klein07,hirayama08,shaji08}

On the other hand, it is known that the multi-valley structure of the conduction band in Si may pose a challenge for quantum computing.\cite{koiller02}  In the context of quantum dot spin qubits in strained silicon heterostructures, it is necessary for the two-fold degeneracy of the two-dimensional electron gas to be lifted sufficiently that spins, not valleys, form the relevant two-level system.\cite{eriksson04}  

The magnitude of the valley splitting has now been measured in strained Si quantum wells at both high fields \cite{weitz96,koester97,khrapai03,lai04} and low fields.\cite{goswami07}  In certain cases, the measured values are 1-2 orders of magnitude smaller than expected from theoretical calculations.\cite{boykin04,boykin04b,nestoklon06}  Such splitting is too small for robust quantum computing applications. It appears that the unexpected suppression is caused by atomic-scale steps at the quantum well interface, as originally suggested by Ando.\cite{ando79}  In this picture, the steps can arise from the underlying miscut in the growth substrate, which is transferred to the interface by conformal epitaxial deposition.  Alternatively, they can be a consequence of strain-induced roughness.  In either case, when an electronic wavefunction extends laterally over many steps, the valley splitting cannot be maximized on every step, leading to an overall suppression of the valley splitting.  Since the valleys states include a phase factor whose value depends on the position of the interface,\cite{friesen07} the suppression of valley splitting can be interpreted as an interference effect caused by multiple steps.\cite{friesen06}  

It should therefore be possible to overcome the suppression of valley splitting by laterally confining the wavefunction to a small number of steps.  This proposal was put to the test in recent experiments by probing the valley splitting under strong electrostatic or magnetic confinement, in a quantum point contact geometry.\cite{goswami07,mcguire08}  Under such conditions, the valley splitting was restored to its expected theoretical value.  By the same token, a smooth interface could partially explain the exceptionally large valley splitting observed at a Si/SiO$_2$ inversion layer.\cite{takashina06}

For quantum computing applications, we are most interested in quantum dot devices, where the confinement is controlled by the top-gate geometry and the application of gate voltages and magnetic fields.  Recent experiments have shown that excitation energies in few-electron quantum dots may be of order 0.1-0.3~meV, which provides an estimate for the valley splitting.\cite{shaji08}  From a quantum computing perspective, the size of this splitting is encouraging, since it is larger than both the typical electron temperature in a dilution refrigerator ($\sim 150$~mK), and the Zeeman spin splitting in fields up to about 2~T. (Note that valley splitting and Zeeman splitting are both functions of the magnetic field.)  The magnitude of the valley splitting therefore appears sufficient for quantum computing in many cases.  However, it is essential to understand the complicated dependence of valley splitting on the shape of the quantum dot and on the materials parameters associated with silicon heterostructures.  In the context of decoherence, it is important to understand how valley-orbit coupling causes quantum dot orbitals to hybridize.  The latter effect plays a role in valley scattering, and can allow the spin qubit to leak into a higher dimensional spin-valley Hilbert space.  

There are many possible approaches to modeling silicon quantum dots, ranging from highly simplified geometries, such as circular quantum dots on a flat quantum well interface, to realistic three-dimensional structures with top-gates and disordered interfaces.  While the simple geometries may be treated analytically, the more realistic geometries require numerical methods.  In this paper, we take a middle road by solving a somewhat complicated model that still allows analytical solutions.  Specifically, we study inter-orbital valley coupling in a quantum dot with an elliptical shape and arbitrary step orientation, in parallel or perpendicular magnetic fields.  In a previous numerical analysis, we also included the effects of interfacial disorder in circular dots,\cite{friesen06}  although we did not account for inter-orbital valley coupling.  In that case, we observed  that interfacial step disorder has a strong moderating effect on the suppression of valley splitting.  Because disorder is not easy to accommodate in our analytical treatment, we focus here on a uniform interface.  However, we do allow for an \textit{effective} tilt angle, which approximates some aspects of disorder.

The paper is outlined as follows.  In Sec.~\ref{sec:VOC} we describe our theoretical method and procedure.  In Sec.~\ref{sec:parallel}, we consider a quantum dot geometry with the magnetic field oriented parallel to the sample surface.  General solutions are obtained for the valley coupling in an elliptical quantum dot, whose orientation with respect to the steps is arbitrary.  We solve several important limiting cases, including high aspect ratios and circular quantum dots.  In Sec.~\ref{sec:perpendicular}, we study circular and elliptical quantum dots in a perpendicular field geometry.  In Sec.~\ref{sec:discussion}, we summarize our main results and provide an intuitive explanation for the origin of valley-orbit coupling.  We discuss some consequences of valley-orbit coupling and why valley splitting is not an entirely well-defined quantity.  We also discuss materials parameters that affect the magnitude of the valley splitting, such as disorder at the interface, and we suggest future experiments to characterize these materials properties.  In the appendix, we provide a discussion of the important approximations used in this paper, including the treatment of the the interfacial tilt as smooth and uniform.

\section{Valley-orbit coupling} \label{sec:VOC}
\subsection{Effective mass formalism}\label{sec:VOCA}
We build on the work of Ref.~\onlinecite{friesen07} to develop an effective mass theory of valley coupling in a quantum dot.  In contrast with most previous theories, we now include inter-orbital couplings.  These have a significant effect on the valley splitting in typical qubit devices, and they are essential for understanding valley scattering.

In this theory, we treat the valley coupling as a perturbation.
Thus, at zeroth order, valleys do not enter the analysis except through the anisotropic effective mass tensor.  The resulting equations have a simple form for inversion layers\cite{ando82} and strained silicon heterostructures,\cite{schaeffler97} since only $z$ valleys are low enough in energy to play a role in the calculations.  In this case, there is only one envelope function equation:  the effective zeroth order Hamiltonian,
\begin{eqnarray}  & & \hspace{-.2in}
H_0=\sum_{j=x,y,z} \frac{1}{2m_j} \left[ -i\hbar \frac{\partial}{\partial r'_j} 
+e A_j(\mathbf{r}') \right]^2 
\label{eq:schrorot0}
\\ & & \hspace{0.9in} 
+ V_\text{QW}(z') + V_\text{QD}(x',y')  . \nonumber
\end{eqnarray}

In Eq.~(\ref{eq:schrorot0}), we refer to two different coordinate systems.  The unprimed coordinate system $(x,y,z)$ is aligned with the crystallographic axes, while the primed coordinate system $(x',y',z')$ is aligned with the growth axis, where $\hat{z}'$ is perpendicular to the plane of the quantum well.  For the $z$ valleys, the transverse effective mass is given by $m_x=m_y=m_t\simeq 0.19m_0$, while the longitudinal effective mass is given by $m_z=m_l\simeq 0.92m_0$.  The confinement potentials include $V_\text{QW}$, the vertical quantum well potential, and $V_\text{QD}$, the lateral quantum dot potential.  The eigenstates of $H_0$
are denoted as envelope functions, $F_n(\mathbf{r}')$, with the discrete orbital index $n$.  
Since $H_0$ does not depend on the individual $z$ valleys, the energy eigenvalues $\varepsilon_n$ are doubly-degenerate.  

At first order in the perturbation theory, we introduce a valley coupling potential $V_v$.  Following 
Fritzsche \cite{fritzsche} and Twose,\cite{twose} we can obtain a set of coupled equations for
the two-valley system:
\begin{equation}
\sum_n \sum_{v=\pm 1} \alpha_{\tilde{n},n,v} 
e^{ivk_0z}[H_0+V_v(\mathbf{r}')-E_{\tilde{n}}]
F_{n}(\mathbf{r}')=0 .  \label{eq:envelope}
\end{equation}
Here, the index $v=\pm 1$ refers to the valley centered at ${\bf k}_v=vk_0 \hat{\bf z}$, where $k_0\simeq 0.82 (2\pi/a)$ and $a=5.43$~\AA~is the Si cubic unit cell dimension.
The interaction $V_v$ mixes unperturbed orbitals with different quantum numbers $n$, so  Eq.~(\ref{eq:envelope}) must include a sum over the index $n$.
The resulting eigenstates of the perturbation Hamiltonian are indexed by a new label, $\tilde{n}$, with corresponding eigenvalues $E_{\tilde{n}}$ and
eigenvectors ${\bm \alpha}_{\tilde{n}}$, whose components span
the unperturbed basis set $\{ |n \rangle \otimes |v\rangle \}$.  Due to valley-orbit mixing, it is not appropriate to distinguish between orbital and valley quantum numbers in realistic quantum dots, except in certain limiting cases.  The index $\tilde{n}$ is therefore a combined valley-orbital label.

We note that Umklapp terms are absent in the summation of Eq.~(\ref{eq:envelope}).  In the context of valley coupling by a shallow donor, the omission of Umklapp processes is known to give numerical errors in the calculation of binding energies.\cite{altarelli77,pantelides78}  While these might be resolved by going beyond conventional effective mass theory,\cite{resca82} the conventional theory does provide acceptable results and intuitive explanations that are in agreement with more detailed analyses, both for single-valley and multi-valley semiconductors.\cite{kohn}  For quantum wells, the two-valley treatment of Eq.~(\ref{eq:envelope}) provides remarkable agreement with more sophisticated theories.\cite{boykin04,friesen07}  This is not surprising -- one would expect the Umklapp errors to be suppressed in quantum wells, as compared to donors, because the fourier transformed potential $V_\text{QW}(k)$ decays more quickly in reciprocal space.  Based on the overall agreement between effective mass and more detailed theories, we will not pursue Umklapp processes here.  

Each term in Eq.~(\ref{eq:envelope}) corresponds to a different unperturbed valley orbital, identified by indices $v$ and $n$:
\begin{equation}
\langle \mathbf{r}'|n,v \rangle =  e^{ik_0vz}F_{n} (\mathbf{r}') . \label{eq:unperturbedEMA}
\end{equation}
It is also convenient to define a special index referring to the opposite valley, $\bar{v}\equiv -v$, such that
\begin{equation}
\langle \mathbf{r}'|n,\bar{v} \rangle 
=e^{-ik_0vz}F_{n} (\mathbf{r}') .
\end{equation}

As discussed elsewhere,\cite{goswami07,friesen07,friesen06} the valley coupling 
potential takes the form of a $\delta$-function in the effective mass theory:
\begin{equation} \label{eq:dfn1}
V_v(\mathbf{r}') = v_v\delta(z'-z_0),
\end{equation}
where $z'=z_0$ is the position of the sharp quantum well interface. 
In principle, one $\delta$-function potential should be included for each interface of the quantum well.  However, in a typical modulation-doped device, the envelope function is centered near the top interface (closest to the doping layer), and is exponentially suppressed at the bottom interface.  Therefore, we include only one $\delta$-function, for the top interface.  (If necessary, a second interface could easily be included in the theory.)

Equation~(\ref{eq:dfn1}) describes a geometry where the interface
occurs on a horizontal plane in the $(x',y',z')$ coordinate system.  This assumes that the stepped interface can be approximated as uniformly tilted.   The validity of such an approximation is discussed in detail in the Appendix.  In brief, since a single-atom step of height $a/4 = 1.4$~\AA~is much smaller than any effective mass length scale, the continuum approximation is well justified in most situations.  Additionally, the assumption of a smooth interface is well-justified, locally, in the absence of significant step bunching.  

When we compute 
matrix elements, as described below, it is more convenient to transform to 
the crystallographic $(x,y,z)$ coordinate system:
\begin{equation} \label{eq:dfn2}
V_v(\mathbf{r})= v_v\delta(z-z_i(x,y)) ,
\end{equation}
where
\begin{eqnarray}
z_i(x,y) &=& z_0-x \tan \vartheta\cos \varphi-   y \tan \vartheta \sin \varphi
\nonumber \\ 
&\simeq & z_0-x\vartheta \cos \varphi -y\vartheta \sin \varphi . \label{eq:varphi}
\end{eqnarray}
Here, $\vartheta$ is the
interfacial tilt angle, which is typically small.  For example, samples grown on commercial substrates may have $\vartheta \leq 2^\circ$, corresponding to a step width of 3.9~nm.  $\varphi$ is the tilt direction in the $x$-$y$ plane, with $\varphi =0$ corresponding to a downhill slope in the $x$ direction.  
For a quantum dot covering many steps, the valley coupling should depend very weakly on the relative position, $z_0$, as we have verified numerically.  In the opposite limit of very few steps, we should treat the steps explicitly, as described in the appendix.  For the present analysis, where we treat the interface as smoothly tilted, it is appropriate to choose the value of $z_0$ to simplify our calculations.

The parameter $\vartheta$ should be viewed as an effective tilt angle.  (See appendix.)  Normal growth dynamics tend to produce interfaces that are disordered,\cite{lagally} so the local tilt angle can differ from the global tilt.  As discussed below, lower ground state energies can be attained at flat interfaces.  Confined electrons will therefore adjust their center positions slightly, to seek out locally flat regions.  A more careful treatment of this phenomenon, and disorder in general, lies outside the scope of the present work.  However, we can capture some of the relevant physics simply by treating $\vartheta$ as an effective, locally-averaged tilt angle.

To solve Eq.~(\ref{eq:envelope}), we proceed as in Ref.~\onlinecite{friesen07}:  we left-multiply by 
$e^{-iv'k_0z}F_{n'}^*(\mathbf{r}')$, integrate over all space, and remove any resulting small terms.
In particular, we drop terms in the integrand containing fast oscillations.\cite{noteosc}  In this way, we obtain the following expression for the matrix elements of the perturbation Hamiltonian:
\begin{eqnarray} \label{eq:ME1}
&& \langle n',v'|H|n,v\rangle
\\ \nonumber && \hspace{.2in} =
\int e^{i(v-v')k_0z}F_{n'}^*(\mathbf{r}') 
H\, F_{n}(\mathbf{r}') d\mathbf{r}' ,
\end{eqnarray}
where we define $H=H_0+V_v$.

After solving for the valley orbitals, we may finally combine the atomic and confinement scale components of the wavefunction, following the prescription given by Kohn:\cite{kohn}
\begin{eqnarray}
\Psi_{\tilde{n}} (\mathbf{r}) &\simeq& \sum_{n} \left[
\alpha_{\tilde{n},n,-1}e^{-ik_0z}u_{-k_0}(\mathbf{r})
\label{eq:Kohn}
\right. \\ \nonumber && \hspace{.3in} \left.
+\alpha_{\tilde{n},n,+1}e^{+ik_0z}u_{+k_0}(\mathbf{r})
\right] F_{n} (\mathbf{r}') .
\end{eqnarray}
Here, $e^{\pm ik_0z}u_{\pm k_0}(\mathbf{r})$ are the bulk Bloch functions for the two valley minima.  Note that the periodic functions $u_{\pm k_0}(\mathbf{r})$ do not play a role in our calculations of the valley orbitals.  The atomic scale details of the valley coupling are captured in the coupling parameter $v_v$.

\subsection{Selection rules and broken symmetries}\label{sec:selection}
We evaluate the matrix elements of Eq.~(\ref{eq:ME1}) as follows.
For the case $v'=v$, we find
\begin{eqnarray} \label{eq:MEdiag0}
&& \langle n',v|H|n,v\rangle =
\\ \nonumber && \hspace{0.6in}
\delta_{n,n'} \left[ \varepsilon_n
+v_v\int |F_{n}(x',y',z_0)|^2 dx'dy'  \right],
\end{eqnarray}
where we have used Eq.~(\ref{eq:dfn1}).  For every quantum dot geometry studied in this paper, we observe the following form of separability for the envelope function:
\begin{equation}
|F_n(x',y',z')| \simeq |f_n(x',y')|\zeta(z') .
\end{equation}
Eq.~(\ref{eq:MEdiag0}) then reduces to
\begin{equation} \label{eq:MEdiag}
\langle n',v|H|n,v\rangle = \delta_{n,n'} \left[ \varepsilon_n
+v_v\zeta^2(z_0)   \right] ,
\end{equation}
when the envelope functions are properly normalized.

The case $v'=\bar{v}$ can be expressed as
\begin{eqnarray}
&& \langle n',\bar{v}|H|n,v\rangle = v_v\zeta^2(z_0)e^{2ivk_0z_0}
\label{eq:ME2} \\ \nonumber &&  \hspace{.4in} \times
\int e^{-2ivk_0[x'\vartheta \cos \varphi +y'\vartheta \sin \varphi ]}
\\ \nonumber && \hspace{.8in} \times
f_{n'}^*(x',y') 
f_{n}(x',y') dx'dy' . 
\end{eqnarray}
Here, we have made use of Eq.~(\ref{eq:varphi}), and the fact that $(x,y)\simeq (x',y')$ when $\vartheta \simeq 0$.  Note that we have dropped the terms $\langle n',\bar{v}|H_0|n,v\rangle$ because they contain fast oscillations and are therefore small. 
Since all coordinate variables in Eq.~(\ref{eq:ME2}) are primed, we henceforth drop the prime notation for simplicity, except where noted.

When $v'=\bar{v}$ and $\vartheta =0$, Eq.~(\ref{eq:ME2}) simplifies, due to the orthonormality of the valley orbitals.  The result is proportional to $\delta_{n,n'}$.
However, when $\vartheta \neq 0$, the phase factors inside the integral are non-vanishing, and inter-orbital couplings arise.  In this way, we see that the tilt of the quantum well is directly responsible for the
off-diagonal valley-orbit coupling.  Since roughness is ubiquitous in realistic devices, we must always expect some degree of inter-orbital valley coupling.

The emergence of inter-orbital coupling has further interesting consequences.
First, it causes broken time-reversal symmetry.
When $\vartheta =0$, the perturbation produces only intra-orbital terms; the valley coupling is identical for every orbital, and its phase factor is trivial, since it can be eliminated by setting $z_0=0$.  But when $\vartheta \neq 0$, the inter-orbital couplings are complex, and they cannot be trivially removed.  The perturbation Hamiltonian is intrinsically complex, and time reversal symmetry is broken.\cite{schiff}   Similar behavior is observed in the multi-valley graphene system, with important implications for weak localization.\cite{morozov06,morpurgo06,beenakker08}
Second, the valley index can no longer be treated as a good quantum number.  When $\vartheta =0$, the valley labels are universal, in the sense that they have the same meaning for every orbital.  In a square well, for example, the wavefunctions can be unambiguously characterized as odd or even.\cite{friesen07}  However, when $\vartheta \neq 0$, there is no universal labeling scheme.  Valley-orbit coupling causes level mixing, which differs from orbital to orbital.  

\subsection{Summary of approximations and their consequences}
The effective mass approximation forms the basis of our analysis, as it allows us to analyze quantum dots in terms of their envelope functions rather than their wavefunctions.  The approximation is a natural one, based on the large separation of lengths scales:\cite{bastard}  atomic ($\sim$0.1~nm) vs.\ confinement (10-100~nm).  One could expect the approximation to break down when the confinement potential is sharp, since the two length scales are then comparable.  However, the effective mass approximation remains viable for many materials, including silicon, even near an abrupt interface.\cite{ando82}  This is because the valleys are roughly parabolic in the regions of the Brillouin zone where the wavefunction is concentrated.

Valley coupling is therefore a higher order effect in the effective mass approximation.  A confinement potential with components at large $k$ will couple the valleys in silicon, leading to an envelope function equation of the form of Eq.~(\ref{eq:envelope}), where the leading order approximation for the coupling takes the form of a $\delta$-function, as in Eq.~(\ref{eq:dfn2}).  When the interface is aligned with the crystallographic axes, the interface position $z_i$ is a constant.  This leads to our first selection rule, that a sharp interface always causes inter-valley coupling within a given orbital.  We note however, that even if the effective mass approximation were to break down near a sharp interface, the valley degeneracy would still be lifted, as confirmed by atomistic theories.\cite{boykin04}  Comparison between atomistic and effective mass theories generally confirms the effective mass approach.\cite{nestoklon06,friesen07}  In the absence of any interfacial tilt, the valley coupling is strictly intra-orbital.  This is a consequence of the assumed separation of variables in the confinement potential, which forms an excellent approximation in quasi-2D devices.\cite{daviesbook}

When the interface is not aligned with the crystallographic axes, we have chosen to treat the interface as smoothly tilted, since single atomic steps are much smaller than effective mass length scales.  The approximation leads immediately to our second selection rule, that a tilted interface causes inter-orbital coupling between opposing valleys, $v$ and $\bar{v}$.  Of course, the smooth interface approximation will not be valid if the misalignment is highly non-uniform, as in the case of multiple step bunching.  In the appendix, we consider several scenarios where the smooth interface approximation breaks down, most notably in cases involving wide steps.  For very wide steps, the smooth approximation leads to a very weak valley coupling.  A more accurate treatment, taking into account the relaxation mechanism described in appendix~A.1, would suppress the valley coupling even beyond the predictions of the present theory.

\section{Parallel Field Geometry} \label{sec:parallel}
We now analyze some specific geometries.  We first consider the case where the magnetic field is
oriented parallel to the sample.  Because the $x$ and $y$ directions
are now inequivalent, even in the absence of interfacial steps, the angular momentum quantum
number $l$ is not a good quantum number.  

We initially treat the problem as an
anisotropic 3D simple harmonic oscillator, with lateral confinement frequencies $\omega_x$
and $\omega_y$.  In the most general case, we assume the oscillator frequencies are unequal:  $\omega_x\neq\omega_y$.  
The quantum well confinement potential is also taken to be
parabolic, with $\omega_z \gg \omega_x,\omega_y$.  Because of the strong confinement along $\hat{\bf z}$, the quantum dot wavefunction is approximately separable in the variable $z$, despite the presence of a magnetic field.  However, we will find that the wavefunction acquires an intrinsic phase when the interface is tilted, due to valley coupling.  Since the magnetic field also produces a non-trivial phase, which could potentially lead to interference effects, it is important to work through the details of the problem without assuming separability, \textit{a priori}.

We take the magnetic field to be oriented along one of the major elliptical axes ($\hat{\bf y}$), otherwise the problem becomes intractable.  Adopting the symmetric gauge
$\mathbf{A}=(z,0,-x)B/2$ for the vector potential, the envelope Hamiltonian becomes
\begin{equation}
H_0=\sum_{j=x,y,z} \left\{ \frac{1}{2m_j} \left[-i\hbar \frac{\partial}{\partial x_j}
+eA_j \right]^2 + \frac{m_j\omega_j^2 x_j^2}{2} \right\} . \label{eq:Hxyz}
\end{equation}
The eigenvalue problem is separable in the variable $y$, leading to solutions of the form
\begin{equation}
F_{n_x,n_y}(\mathbf{r})=f_{n_x}(x,z)g_{n_y}(y) ,
\end{equation}
where $n_x$ and $n_y$ are non-negative integers.  The solutions for $g_{n_y}$ are given by
\begin{eqnarray}
g_{n_y}(y) &=& \left( 2^{n_y}n_y! \right)^{-1/2} 
\left( \frac{m_t\omega_y}{\pi\hbar} 
\right)^{1/4} e^{-(m_t\omega_y/2\hbar)y^2 }
\nonumber \\ && \hspace{.2in} \times
H_{n_y}\left(\sqrt{\frac{m_t\omega_y}{\hbar}} y\right) , \label{eq:phiny}
\end{eqnarray}
where $H_n(x)$ is a Hermite polynomial.\cite{stegun}

The magnetic field couples the $x$ and $z$ variables in Eq.~(\ref{eq:Hxyz}), making the
problem somewhat more complicated.  However, by rescaling the coordinate variables as
\begin{eqnarray}
x &=& (m_l/m_t)^{1/4} \tilde{x} , \label{eq:rescal} \\
z &=& (m_t/m_l)^{1/4} \tilde{z} , \label{eq:rescal2}
\end{eqnarray}
the $x$-$z$ Hamiltonian is brought into a more tractable form:
\begin{equation}
\tilde{H}_0=\frac{1}{2\mu} \left[ -i\hbar \tilde{\bm \nabla} +e\tilde{\mathbf{A}} \right]^2
+\frac{\mu}{2} \left( \omega_x^2\tilde{x}^2+\omega_z^2\tilde{z}^2 \right) ,
\label{eq:Htilde}
\end{equation}
where $\mu \equiv \sqrt{m_tm_l}$, and we define a scaled vector potential,
$\tilde{\mathbf{A}} = (\tilde{z},0,-\tilde{x})B/2$.

In principle, the $\omega_z$ oscillator possesses a distinct quantum number, $n_z$.  However,
we restrict our investigation to the first subband approximation ($n_z=0$), since 
$\omega_z\gg\omega_x$.  We also assume that the quantum well confinement is stronger than the magnetic confinement, $\omega_z\gg\omega_c$, as consistent with typical laboratory conditions.  (Here, $\omega_c=e|B|/m_t$ is the cyclotron frequency.)  Although valley coupling can cause mixing of the higher subbands, the relevant energy scales are very large, and the couplings are small.  We ignore such mixing here.

An exact solution for Eq.~(\ref{eq:Htilde}) was first obtained in Ref.~\onlinecite{dippel94}.  For the problem considered here, the results can be expressed as
\begin{widetext}
\begin{eqnarray}
f_{n_x}(x,z) &=& N_{n_x} \exp \left\{ \frac{iJ xz}{\hbar} 
-\frac{M_1\omega_1(m_l/m_t)^{1/2}z^2+2iL M_1\omega_1M_2\omega_2xz+M_2\omega_2(m_t/m_l)^{1/2}x^2}
{2\hbar (L^2M_1\omega_1M_2\omega_2+1)} \right\}
\\ \nonumber && \hspace{1in} \times
\sum_{l=0}^{n_x} \frac{c_{l} n!}{l!(n_x-l)!} H_{n_x-l} \left[
\frac{\sqrt{2M_2\omega_2/\hbar}(L M_1\omega_1(m_l/m_t)^{1/4}z-i(m_t/m_l)^{1/4}x)}
{L^2M_1\omega_1M_2\omega_2+1} \right] ,
\end{eqnarray}
where
\end{widetext}
\begin{eqnarray}
J &=& -\frac{\mu}{2} \left( \frac{\omega_z^2-\omega_x^2}{\omega_c} \right)
\\ && \nonumber
+\frac{\mu}{2\omega_c}
\sqrt{(\omega_x^2+\omega_z^2+\omega_c^2)^2-4\omega_x^2\omega_z^2} , 
\end{eqnarray}
\begin{equation}
L = \frac{\omega_c}{\mu 
\sqrt{(\omega_x^2+\omega_z^2+\omega_c^2)^2-4\omega_x^2\omega_z^2}} , 
\end{equation}
\begin{eqnarray}
M_{1,2} &=& 
\\ \nonumber && \hspace{-.3in}
\frac{2\mu \sqrt{(\omega_x^2+\omega_z^2+\omega_c^2)^2-4\omega_x^2\omega_z^2}}
{(\omega_z^2-\omega_x^2\pm \omega_c^2)+
\sqrt{(\omega_x^2+\omega_z^2+\omega_c^2)^2-4\omega_x^2\omega_z^2}} , 
\end{eqnarray}
\begin{eqnarray}
\omega_{1,2} &=& \frac{1}{\sqrt{2}}
\bigglb[ \omega_x^2+\omega_z^2+\omega_c^2
\\ \nonumber && \hspace{.3in}
\pm \sqrt{(\omega_x^2+\omega_z^2+\omega_c^2)^2-4\omega_x^2\omega_z^2} \biggrb]^{1/2} , 
\end{eqnarray}
\begin{equation}
N_{n_x} = \left(\frac{-i}{2} \right)^{n_x} \left[
\frac{\sqrt{M_1\omega_1M_2\omega_2}}{\pi^2\hbar (n_x)!(L^2M_1\omega_1M_2\omega_2+1) }
\right]^{1/2} \hspace{-.1in} , 
\end{equation}
\begin{equation}
c_{l} = \left\{ \begin{array}{ll}
(4G)^{l/2}\Gamma \left(\frac{l+1}{2} \right) 
 & \text{($l$ even),} \\
0 &  \text{($l$ odd),} \end{array} \right. 
\end{equation}
and
\begin{equation}
G = \frac{4}{L^2M_1\omega_1M_2\omega_2+1}-1 .
\end{equation}
Here, $\Gamma (p)$ is the Gamma function.\cite{stegun}
The energy eigenvalues corresponding to $F_{n_x,n_y}$ are 
\begin{equation}
\varepsilon_{n_x,n_y}=\frac{\hbar\omega_1}{2}+\left(n_x+\frac{1}{2} \right)\hbar\omega_2
+\left(n_y+\frac{1}{2} \right)\hbar\omega_y .
\end{equation}

We now apply the limit $\omega_z\gg\omega_x,\omega_c$, obtaining 
\begin{eqnarray}
\omega_1 &=& \omega_z +\frac{\omega_c^2}{2\omega_z} + \mathcal{O}[\omega_z^{-3}] \\
\omega_2 &=& \omega_x + \mathcal{O} [\omega_z^{-2}] .
\end{eqnarray}
Including the leading order correction in $(\omega_c/\omega_z)$, the quantum dot energy is then given by
\begin{eqnarray}
\varepsilon_{n_x,n_y} &\simeq & \frac{\hbar}{2}\left(\omega_z+\frac{\omega_c^2}{2\omega_z} \right)
\label{eq:Epar} \\ \nonumber && \hspace{0in}
+\left(n_x+\frac{1}{2} \right)\hbar\omega_x
+\left(n_y+\frac{1}{2} \right)\hbar\omega_y ,
\end{eqnarray}
with the corresponding $x$-$z$ eigenfunction
\begin{eqnarray} \label{eq:psi2}
f_{n_x}(x,z) &\simeq & 
\left( \frac{\sqrt{m_tm_l\omega_x\omega_z}}{\pi\hbar 2^{n_x}(n_x)!} \right)^{1/2}
\label{eq:psiell} \\ \nonumber && \hspace{-.5in} \times
\exp \left\{ -\frac{m_t\omega_x}{2\hbar}x^2+\frac{i\sqrt{m_tm_l}\omega_c}{2\hbar}xz
-\frac{m_l\omega_z}{2\hbar}z^2 \right\}
\\ \nonumber && \hspace{-.5in} \times
H_{n_x}\left(\sqrt{\frac{m_t\omega_x}{\hbar}} x\right) .
\end{eqnarray}
Here, we have made use of a Hermite polynomial identity:
\begin{equation}
2^{n/2}H_n\left( \frac{x+y}{\sqrt{2}} \right) = \sum_{k=0}^n \frac{n!}{k!(n-k)!}H_k(x)H_{n-k}(y) .
\end{equation}
We find that the main effect of the magnetic field is 
to introduce a phase factor into the otherwise separable anisotropic oscillator
solutions.  This phase plays no role when it is included in valley coupling calculations, since it is independent of $n_x$.  The conventional wisdom that quantum dot wavefunctions are separable in the parallel field geometry is therefore also valid in the context of valley coupling.

\subsection{Flat Quantum Well}
We now compute the matrix elements $\langle n_x'n_y'v'|H|n_xn_yv \rangle$ for a flat 
quantum well, treating the valley coupling potential $V_v$ as 
a perturbation.  When $v'=v$, only diagonal matrix elements survive, as explained in Sec.~\ref{sec:selection}.  Note that we
have chosen the center of the harmonic oscillator to be at the origin.  Defining the top of the quantum well as $z=z_0$, we obtain
\begin{equation}
\langle n_xn_yv|H|n_xn_yv \rangle = \varepsilon_{n_x,n_y}
+ v_v\zeta_0^2(z_0) , \label{eq:flat1} 
\end{equation}
with
\begin{equation}
\zeta_0(z_0)=(m_l\omega_z/\pi\hbar)^{1/4} e^{-(m_l\omega_z/2\hbar)z_0^2}  .
\label{eq:zeta0}
\end{equation}
This final expression was obtained by treating the vertical confinement potential as harmonic.

The approximate separability of variables in Eq.~(\ref{eq:psiell}) suggests that  $\hbar \omega_z$ and $\zeta_0$ could be replaced by more physically motivated quantities, such as eigenstates of a finite triangular well.  Indeed, throughout this work, we will view $v_v\zeta_0^2$ as a scaling factor for the valley coupling, whose value can be measured experimentally and inserted, phenomenologically, into the theoretical expressions.  When microscopic disorder is present at the quantum well interface and significant details about the valley coupling are not known precisely, such an approach becomes quite practical.  Consistent with this view, we henceforth drop the argument $z_0$ in $\zeta_0$.  

\begin{figure}[t] 
  \centering
  \includegraphics[width=2.6in,keepaspectratio]{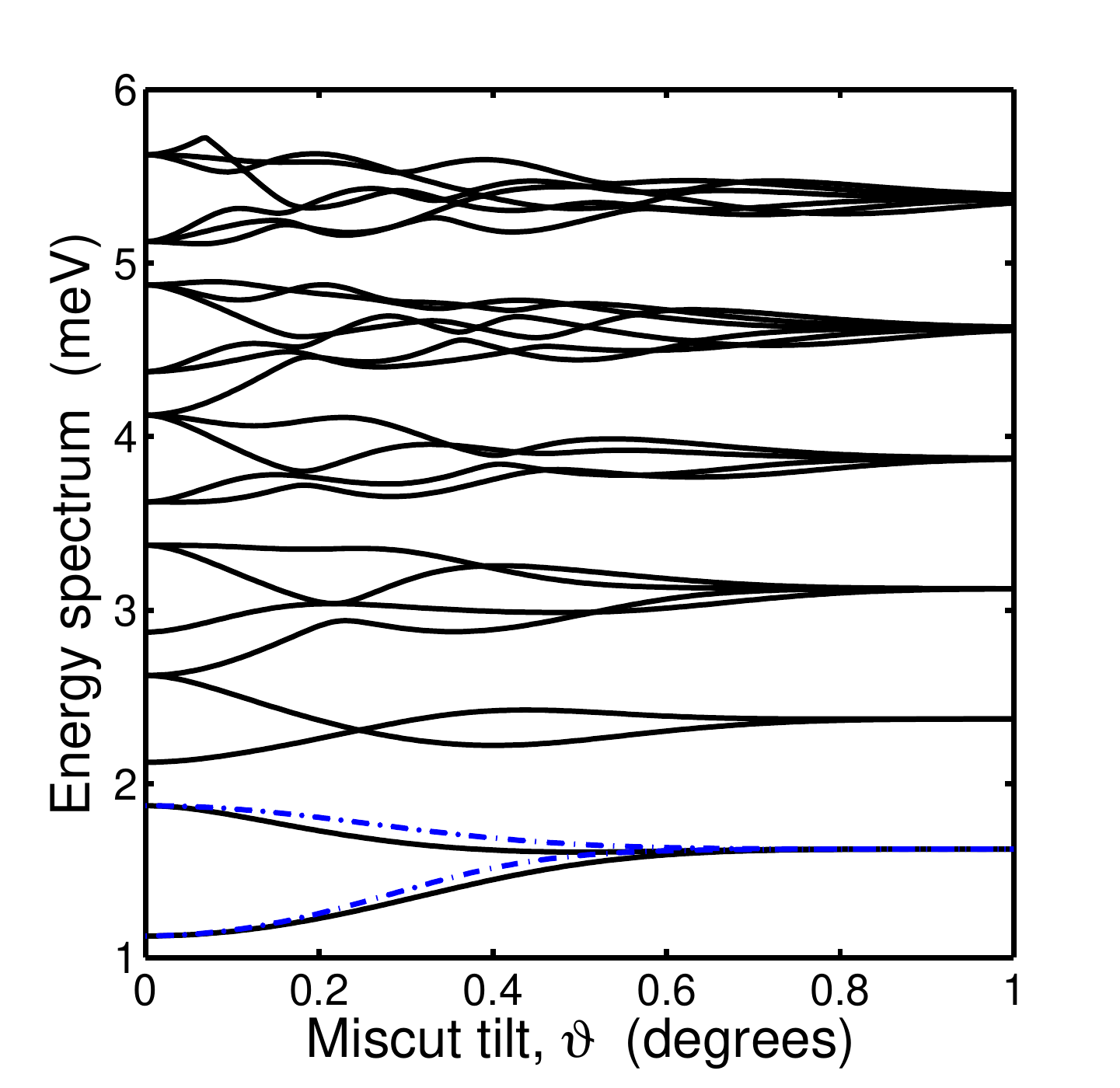}
  \caption{(Color online) Energy levels for an elliptical quantum dot in the parallel field geometry, including valley coupling.  Dot parameters are appropriate for a single-electron spin qubit:  $\hbar \omega_x=1.5$~meV, $\hbar \omega_y=0.75$~meV, $v_v\zeta_0^2=0.5$~meV.  Uniform, lateral step orientation is set to $\varphi =\pi/4$.  Blue dot-dashed lines show the analytical approximation, obtained by restricting the analysis to the lowest two orbital levels in Eq.~(\ref{eq:fourfour}).  The subband energy shift (a constant) has been dropped here, for simplicity.}
  \label{fig:elltilt2}
\end{figure}

For the off-diagonal matrix elements $\langle n_x'n_y'\bar{v}|H|n_xn_yv \rangle$, the orthogonality of the eigenfunctions restricts the off-diagonal couplings to cases where $n_x=n_x'$ and $n_y=n_y'$, as discussed in Sec.~\ref{sec:selection}.  (Recall that $\vartheta =0$ in this section.)  This gives the trivial result
\begin{equation}
\langle n_x'n_y'\bar{v}|H|n_xn_yv \rangle = \delta_{n_x,n_x'}\delta_{n_y,n_y'}
v_v\zeta_0^2 e^{2ik_0z_0v} .
\end{equation}
The two-fold degeneracy of each orbital level is lifted by a fixed amount, independent of the quantum state.  The ground state gap for each pair of states is given by $E_g=2v_v\zeta_0^2$.  This is the conventional ``valley splitting."

\subsection{Tilted Quantum Well} \label{eq:perptilt}
We turn to the more realistic case of a quantum well tilted downwards at an angle
$\vartheta$, towards the direction $\varphi$ in the $x$-$y$ plane, as described in 
Eq.~(\ref{eq:varphi}).
When $v'=v$, as in Eq.~(\ref{eq:MEdiag0}), then  to leading order in the small parameter $(\omega_c/\omega_z)$, the Hamiltonian matrix elements are real and exclusively diagonal.  They are
identical to their counterparts for a flat quantum well.
However when $v'=\bar{v}$, the valley interaction induces nontrivial
couplings between the quantum dot levels, 
leading to new avoided crossings in the energy spectrum, and to fundamentally new physics.

The coupling matrix elements are solved using an explicit expansion for the Hermite polynomials,
\begin{equation}
H_n(x)=\sum_{j=0}^{\lfloor n/2 \rfloor} \frac{(-1)^jn!}{j!(n-2j)!}(2x)^{n-2j} \,,
\end{equation}
where the floor function $\lfloor m\rfloor$ is defined as the greatest integer less than
or equal to $m$, and $n!$ is the factorial function with $0!=1$.  
We obtain
\begin{widetext}
\begin{eqnarray} \label{eq:bigH}
&& \langle n_x'n_y'\bar{v}|H|n_xn_yv\rangle =
v_v\zeta_0^2 e^{2ik_0z_0v}
\left[ \frac{m_t^2\omega_x\omega_y}
{\pi^2\hbar^22^{n_x+n_x'+n_y+n_y'} (n_x)!(n_x')!(n_y)!(n_y')!} \right]^{1/2}
\\ \nonumber && \hspace{.65in} \times
\int_{-\infty}^\infty \! dx dy \,
\exp \left[ -\frac{m_t\omega_x}{\hbar} x^2
-\frac{m_t\omega_y}{\hbar} y^2 \right] \exp \left[
-2ik_0v(x\vartheta \cos \varphi +y\vartheta \sin \varphi) \right]
\\ \nonumber && \hspace{.65in} \times \;
H_{n_x} \left[ \sqrt{\frac{m_t\omega_x}{\hbar}}x \right]
H_{n_x'} \left[ \sqrt{\frac{m_t\omega_x}{\hbar}}x \right]
H_{n_y} \left[ \sqrt{\frac{m_t\omega_y}{\hbar}}y \right]
H_{n_y'} \left[ \sqrt{\frac{m_t\omega_y}{\hbar}}y \right] .
\\ \nonumber && \hspace{.5in} 
= v_v\zeta_0^2 e^{2ik_0z_0v}
\sqrt{(n_x)!(n_x')!(n_y)!(n_y')!} 
\exp \left[ -\frac{\hbar k_0^2\vartheta^2}{m_t}
\left( \frac{ \cos^2 \varphi}{\omega_x}+\frac{ \sin^2 \varphi}{\omega_y} \right) \right]
\left( \frac{v}{i\sqrt{2}} \right)^{n_x+n_x'+n_y+n_y'}
\\ \nonumber && \hspace{.65in} \times
\sum_{l=0}^{\lfloor n_x/2\rfloor} \sum_{l'=0}^{\lfloor n_x'/2\rfloor}
\sum_{j=0}^{\lfloor n_y/2\rfloor} \sum_{j'=0}^{\lfloor n_y'/2\rfloor} 
\left[ l!(l')!(n_x-2l)!(n_x'-2l')!j!(j')!(n_y-2j)!(n_y'-2j')! \right]^{-1}
\\ \nonumber && \hspace{.65in} \times \;
H_{n_x+n_x'-2(l+l')} 
\left[ \sqrt{\frac{\hbar}{m_t\omega_x}}k_0\vartheta \cos \varphi \right]
H_{n_y+n_y'-2(j+j')} 
\left[ \sqrt{\frac{\hbar}{m_t\omega_y}}k_0\vartheta \sin \varphi \right] .
\end{eqnarray}
\end{widetext}
This result is separable with respect to the $x$ and $y$ indices, $\{n_x,n_x'\}$ and $\{n_y,n_y'\}$, indicating that mixing occurs among $x$ orbitals or $y$ orbitals, with no $x$-$y$ cross-coupling.  
At this point, we invoke our freedom to set $z_0=0$, as discussed in Sec.~\ref{sec:VOCA}.

The couplings appearing in Eq.~(\ref{eq:bigH}) suggest a rather complicated dependence of the valley splitting on the tilt angle $\vartheta$.  The energy spectrum for typical dot parameters is shown in Fig.~\ref{fig:elltilt2}.  The different asymptotic behaviors can be understood as follows.  
In the limit $\vartheta \rightarrow 0$, rotational symmetry is recovered (see Sec.~\ref{sec:selection}), and valley coupling does not mix the orbital levels.  In this case, the valley states in each orbital level are split by the same amount.  However, for realistic quantum dot diameters ($\sim 50-100$~nm), the orbital spacing can be smaller than the single-particle valley splitting ($\sim 0.3-0.5$~meV), as in Fig.~\ref{fig:elltilt2}.  In this case, pairs of states with the same orbital quantum numbers are difficult to identify, due to the presence of valleys.

\begin{figure}[t] 
  \centering
  \includegraphics[width=2.5in,keepaspectratio]{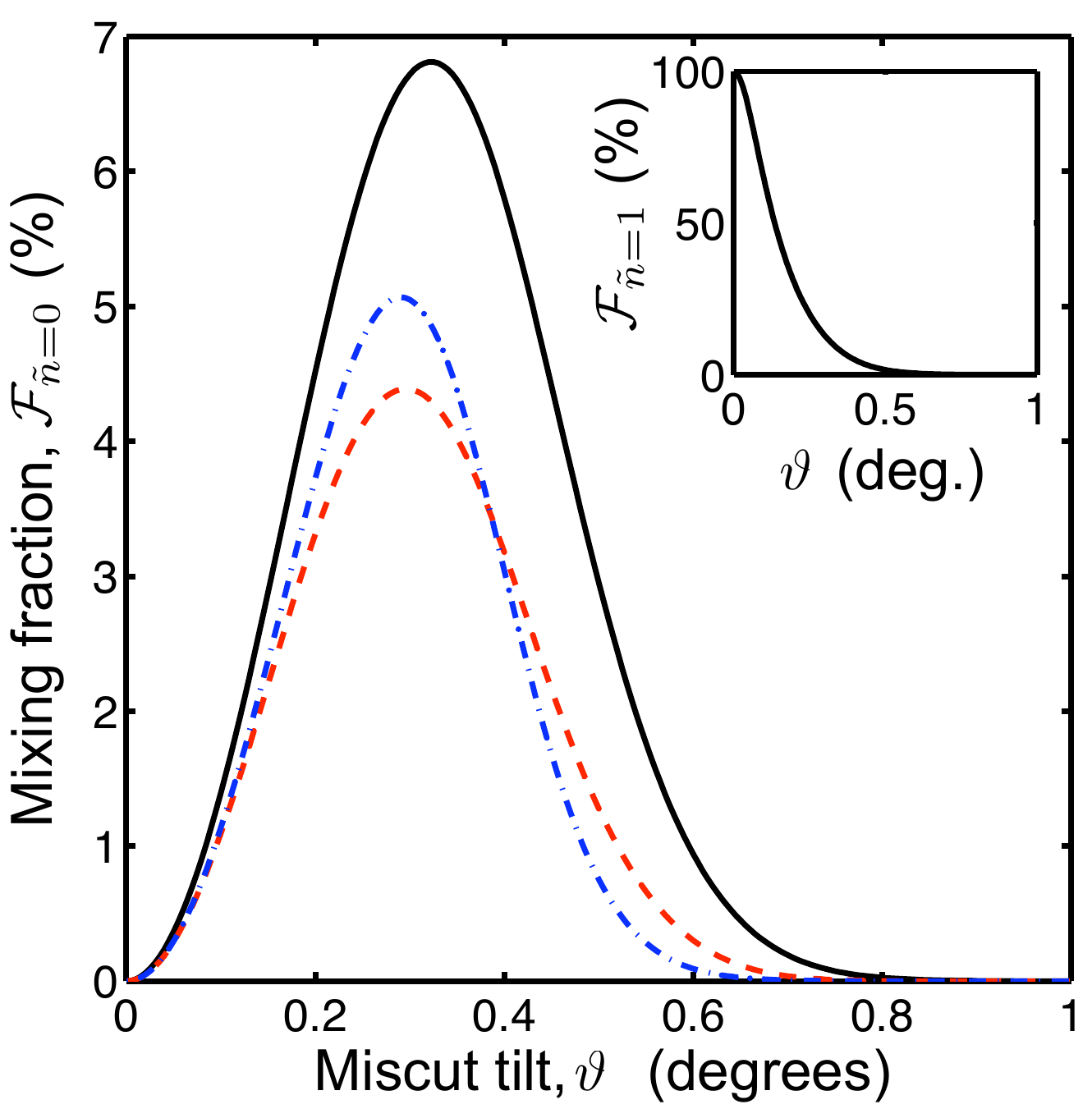}
  \caption{(Color online) Mixing fraction, Eq.~(\ref{eq:mixing}), due to inter-orbital valley coupling.  Results are shown for the same quantum dot parameters used in Fig.~\ref{fig:elltilt2}.  The solid black line is the projection of the perturbed ground state ($\tilde{n}=0$) onto the unperturbed excited states ($n>0$).  The red dashed line shows the main contribution to ${\cal F}_{\tilde{n}=0}$ coming from just the first excited orbital ($n=1$).  Blue dot-dashed line shows the corresponding analytical result of Eq.~(\ref{eq:fourfour}), as obtained within a restricted subspace.  Inset: the projection of the first excited state ($\tilde{n}=1$) onto the unperturbed excited states ($n>0$).}
  \label{fig:VOC}
\end{figure}

When $\vartheta >0$, valley-orbit coupling mixes the unperturbed orbitals, leading to avoided crossings.
The degree of mixing affects the valley scattering.  We can quantify this effect by defining the mixing fraction ${\cal F}_{\tilde{n}=0}$, which describes the projection of the perturbed ground state ($\tilde{n}=0$) onto the unperturbed excited states ($n>0$):  
\begin{equation}
{\cal F}_{\tilde{n}}=\sum_{n>0,v=\pm1} |\alpha_{\tilde{n},n,v}|^2 . \label{eq:mixing}
\end{equation}
Some typical results for the mixing fraction are shown in Fig.~\ref{fig:VOC}.  We see that ${\cal F}_{\tilde{n}=0}$ vanishes when $\vartheta \rightarrow 0$, as explained above, and goes through a maximum value.  In the limit $\vartheta \gg 1$, the valley coupling is suppressed for all orbital states.  (See Fig.~\ref{fig:elltilt2}.)  This is a consequence of destructive valley interference, caused by multiple step coverage.\cite{goswami07,friesen06} The mixing fraction is also suppressed for the same reason.  In Secs.~\ref{sec:orbital}-E, we will obtain analytical expressions for the valley-orbit coupling and the ground state gap, for certain quantum dot geometries.

The first excited state experiences stronger valley-orbit coupling than the ground state, because of its proximity to other nearby energy levels.  This is demonstrated vividly in Fig.~\ref{fig:elltilt2}.  In the two limits $\vartheta =0$ and $\vartheta \gg 1$, the orbital quantum numbers $n_{x,y}$ are good quantum numbers.  When $\vartheta =0$, the valley splitting is larger than the orbital spacing, so $n_y$ will typically vary on successive energy levels (in this case, $n_y=0,1,0,\dots$, from bottom to top).  However, when $\vartheta \gg 1$, the lowest pair of eigenstates becomes nearly degenerate, with the same quantum number, $n_y=0$.  Thus, the orbital character of the first excited state changes completely, from $n_y=1$ to 0, as a result of valley-orbit coupling.
In the intermediate region, $\vartheta \lesssim 1$, valley-orbit coupling causes a complete mixing of the unperturbed orbitals.  The orbital crossover can also be observed in the inset of Fig.~\ref{fig:VOC}, which shows the mixing fraction for the first excited state.  Clearly we cannot obtain exact expressions for the ground state gap or the mixing fraction.  However, approximations are available in certain limiting cases, as described in the following sections.

\subsection{Limit: Small quantum dots} \label{sec:orbital}
This limit corresponds to the case where orbital splittings are much larger than the valley coupling, leading to particularly simple approximations for the ground state gap.  To leading order, we can treat the ground state by simply ignoring inter-orbital valley couplings.  The quantum numbers $n_x=n_y=0$ then remain good quantum numbers, even after applying the perturbation.  

We can express the Hamiltonian in the restricted manifold $\{ |00+\rangle,|00-\rangle\}$, where $+$ and $-$ refer to the unperturbed valley quantum numbers.  From Eq.~(\ref{eq:bigH}) and Sec.~\ref{sec:selection}, we obtain
\begin{equation}
H\simeq \begin{pmatrix}
\varepsilon_0+\Delta_0 & \Delta_0 \eta \\
\Delta_0 \eta & \varepsilon_0+\Delta_0
\end{pmatrix} , \label{eq:twotwo}
\end{equation}
where we define
\begin{eqnarray}
\varepsilon_0 &=& \hbar (\omega_x+\omega_y)/2 ,\\
\Delta_0 &=& v_v\zeta_0^2 , \label{eq:D0} \\
\eta &=& \exp \left[ -\frac{\hbar k_0^2\vartheta^2}{m_t}
\left( \frac{ \cos^2 \varphi}{\omega_x}+\frac{ \sin^2 \varphi}{\omega_y} \right) \right]  .
\end{eqnarray}
Note that the vertical confinement energy $(\hbar/2)(\omega_z+\omega_c^2/2\omega_z$ is a constant here; we set it to zero for simplicity.

Diagonalization immediately gives the ground state gap,
\begin{eqnarray}
E_g & \simeq & 2\Delta_0\eta
\label{eq:orbital} \\ \nonumber
& = & 2v_v\zeta_0^2
\exp \left[ -\frac{\hbar k_0^2\vartheta^2}{m_t}
\left( \frac{ \cos^2 \varphi}{\omega_x}+\frac{ \sin^2 \varphi}{\omega_y} \right) \right] .
\end{eqnarray}
The quantity $2\Delta_0$ can be recognized as the theoretical maximum of the valley splitting, while $2\Delta_0\eta$ corresponds to the renormalized or suppressed valley splitting, which takes into account the steps at the interface.

\begin{figure}[t] 
  \centering
  \includegraphics[width=2.2in,keepaspectratio]{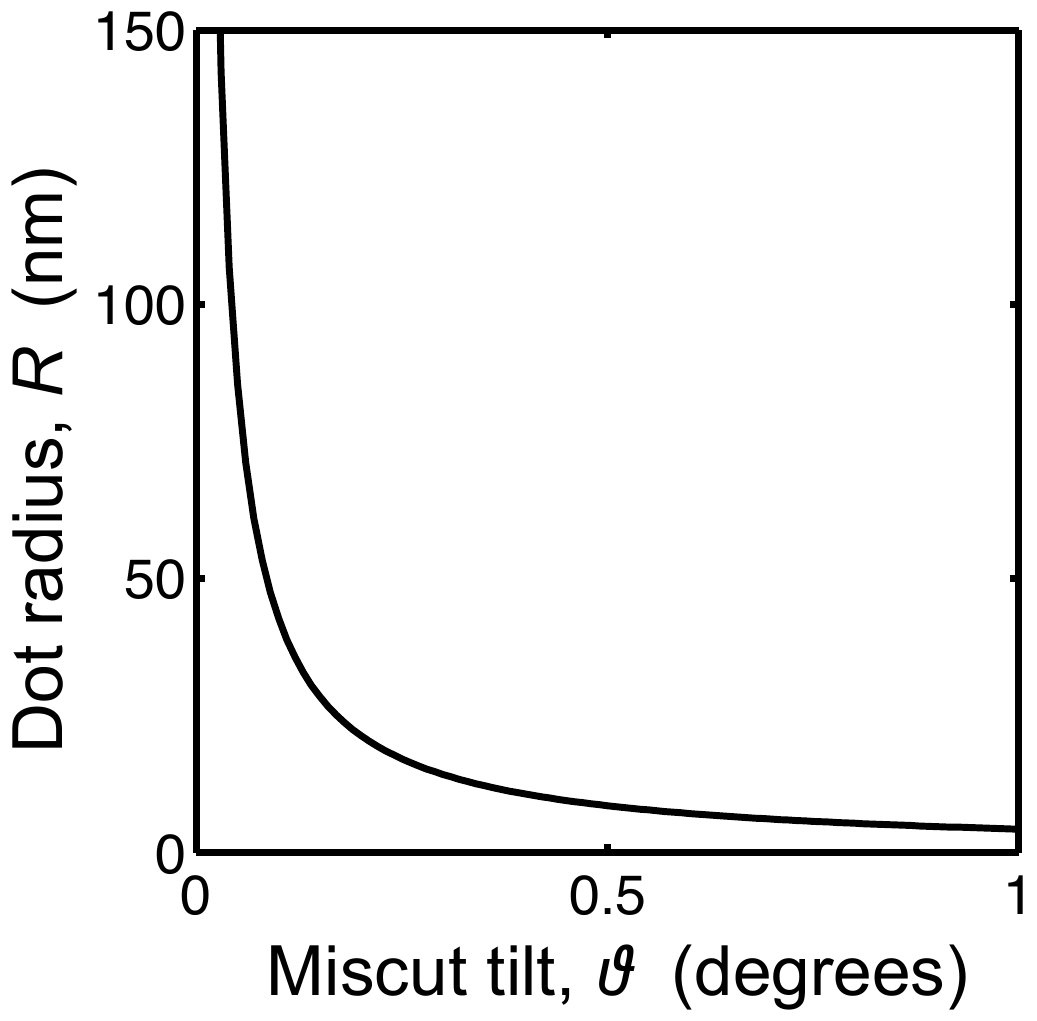}
  \caption{Crossover radius for the suppression of valley coupling caused by a tilted interface.  For a quantum dot lying below the curve, valley splitting is weakly suppressed or unsuppressed.  Above the curve, valley splitting is strongly suppressed.}
  \label{fig:radius}
\end{figure}

The previous approximation is most appropriate for ultra-small quantum dots in a 2DEG , with radii $\ll 20$~nm. Although such devices are challenging to fabricate at the present time, the main result of this section, Eq.~(\ref{eq:orbital}), is more broadly applicable.  This is particularly true when $\vartheta$ is so large that the valley coupling is strongly suppressed.  We can use Eq.~(\ref{eq:orbital}) to determine the crossover angle $\vartheta_s$, above which the valley coupling is exponentially suppressed:
\begin{equation}
\vartheta_s \simeq \sqrt{m_t\omega/\hbar k_0^2} .  
\label{eq:varthetas}
\end{equation}
For a typical quantum dot energy of $\hbar \omega = 0.75$~meV (see Fig.~\ref{fig:elltilt2}), this corresponds to a crossover angle $\vartheta_s=0.26^\circ$.  Taking into account the simple harmonic oscillator radius, $R=\sqrt{\hbar/2m_t\omega}$, we obtain a relation between $R$ and $\vartheta_s$, as shown in Fig.~\ref{fig:radius}.  

\subsection{Limit: High aspect ratio} \label{sec:long}
We now consider slightly larger quantum dots, of interest for quantum computing.  In this regime, the orbital and valley splittings are similar in magnitude.  We further assume that one principle axis of the elliptical potential is much larger than the other ($\omega_x\gg \omega_y$).  Inter-orbital couplings to the ground state are then dominated by a single excited state, $n_y=1$.  We find that this approximation is applicable over a fairly wide range.  For example, Fig.~\ref{fig:VOC} corresponds to a moderate aspect ratio of $\omega_x/\omega_y=2$.  The red dashed line shows results obtained by solving the Hamiltonian with many orbital states, but including just  the contribution from the first excited state.  The results are satisfactory, indicating the accuracy of the approximation.

For the analytical treatment described above, we consider the ordered manifold $\{ |00+\rangle,|00-\rangle,|01+\rangle,|01-\rangle \}$, obtaining
\begin{equation}
H\simeq \begin{pmatrix}
\varepsilon_0+\Delta_0 & \Delta_0 \eta & 0 & i\Delta_0 \eta q \\
\Delta_0 \eta & \varepsilon_0+\Delta_0 & -i\Delta_0 \eta q & 0 \\
0 & i\Delta_0 \eta q & \varepsilon_1+\Delta_0 & \Delta_0 \eta (1-q^2) \\
-i\Delta_0 \eta q & 0 & \Delta_0 \eta (1-q^2) & \varepsilon_1+\Delta_0
\end{pmatrix} . \label{eq:fourfour}
\end{equation}
Here, we use the previous definitions for $\varepsilon_0$, $\Delta_0$, and $\eta$, and we further define
\begin{eqnarray}
\varepsilon_1 &=& \hbar (\omega_x+3\omega_y)/2 ,\\
q &=& (k_0\vartheta \sin \varphi ) \sqrt{2\hbar /m_t\omega_y } .
\end{eqnarray}
As before, the vertical confinement energy is a constant, and we set it to zero.

Diagonalizing $H$ gives the energy spectrum in this restricted manifold.  
Fig.~\ref{fig:elltilt2} shows two two lowest energy levels obtained by this procedure (blue dot-dashed line).  The results are satisfactory, even though the high aspect ratio limit is not well satisfied.  In particular, the ground state gap is found to be quite accurate.  

One main advantage of considering restricted manifolds is that they allow analytic results.  Thus, the ground state gap can be computed, giving
\begin{eqnarray}
E_g &\simeq & \Delta_0\eta q^2
\label{eq:Evaspect} \\ \nonumber && \hspace{-.25in}
+\frac{1}{2}\sqrt{[\hbar \omega_y+\Delta_0 \eta (2-q^2)]^2 +(2\Delta_0 \eta)^2 q^2} 
\\ \nonumber && \hspace{-.25in}
- \frac{1}{2}\sqrt{[\hbar \omega_y-\Delta_0 \eta (2-q^2)]^2 +(2\Delta_0\eta)^2 q^2}  .
\end{eqnarray}
We can also compute the mixing fraction defined in Eq.~(\ref{eq:mixing}): 
\begin{widetext}
\begin{equation}
{\cal F}_{\tilde{n}=0}
= \frac{2(\Delta_0\eta)^2 q^2}{[\hbar \omega_y+\Delta_0 \eta (2-q^2)]^2 +(2\Delta_0\eta)^2 q^2
+[\hbar \omega_y+\Delta_0 \eta (2-q^2)]
\sqrt{[\hbar \omega_y+\Delta_0 \eta (2-q^2)]^2 +(2\Delta_0\eta)^2 q^2}} . \label{eq:Faspect}
\end{equation} 
\end{widetext}
In Fig.~\ref{fig:VOC}, we compare this analytical theory (dotted line) with the full numerical calculation.  Again, we find that the restricted manifold provides a very reasonable approximation.

\subsection{Limit: Circular dot} \label{sec:circle}
Circular quantum dots are of interest for their symmetry properties.  We therefore introduce the angular momentum quantum numbers, $n$ and $l$.  In principle, a parallel magnetic field will break the $x$-$y$ rotational symmetry.  However, for typical laboratory fields, we have shown that the only effect of a parallel field is to add a phase factor to the wavefunction.  The phase factor does not appear in the Hamiltonian matrix elements.  With this caveat, we can ignore the parallel magnetic field in the present analysis, and consider the wavefunction to be symmetric with respect to $x$ and $y$.  

In Sec.~\ref{sec:perpcirc}, below, we solve the circular quantum dot geometry in a perpendicular magnetic field.  According to the previous discussion, we can obtain results appropriate for a parallel field from that analysis, by taking the limit $B\rightarrow 0$.  This amounts to making the substitution
\begin{equation}
b/l_B^2 \rightarrow 2\omega_0m_t/\hbar  \label{eq:Blim}
\end{equation}  
in Eqs.~(\ref{eq:FD})-(\ref{eq:bigME}).
The corresponding energy levels are given by
\begin{eqnarray}
&& \hspace{-.2in} \varepsilon_{nl}=\frac{\hbar \omega_z}{2}+\frac{\hbar\omega_c^2}{4\omega_z}
+\hbar \omega_0 (2n+|l|+1) 
\label{eq:eppar} \\ \nonumber && \hspace{1.5in} \text{(parallel $B$-field)} .
\end{eqnarray}
Here, we have also included the leading contribution to the magnetic confinement energy, in analogy with Eq.~(\ref{eq:Epar}).

The energy spectrum of a circular quantum dot is qualitatively similar to an elliptical dot, and we do not show any plots here.  However, some limiting behaviors are of interest.  In the small dot limit, we evaluate the perturbation Hamiltonian in the restricted manifold $\{|n,l,v\rangle\}=\{|0,0,+\rangle,|0,0,-\rangle\}$, similar  to Sec.~\ref{sec:orbital}.  We obtain the ground state gap
\begin{equation}
E_g\simeq 2\Delta_0\eta ,
\end{equation}
where $\Delta_0$ is the same as before, and we now define
\begin{equation}
\eta =e^{-q^2}\quad\text{with}\quad q=k_0\vartheta\sqrt{\hbar/\omega_0m_t} \,\, .
\label{eq:etapar}
\end{equation}
The result is equivalent to Eq.~(\ref{eq:orbital}) in the isotropic limit $\omega_x=\omega_y=\omega_0$, where we set $\varphi=0$.  

We can also obtain analytical approximations for the circular quantum dot geometry when the orbital and valley splittings are similar in magnitude.  We consider the manifold consisting of the ground state and the lowest excited orbitals.  From Eq.~(\ref{eq:eppar}), we note that $\pm l$ states are degenerate.  The manifold of interest is therefore 6-dimensional, with the quantum numbers $n=0$, $l=0,\pm 1$, and $v=\pm 1$.  In the ordered basis $\{ |0,0,+\rangle,|0,0,-\rangle,|0,1,+\rangle,|0,1,-\rangle,|0,-1,+\rangle,|0,-1,-\rangle \}$, we obtain 
\begin{widetext}
\begin{equation}
H\simeq \begin{pmatrix}
\varepsilon_{0,0}+\Delta_0 & \Delta_0 \eta & 0 & i\Delta_0 \eta q & 0 & i\Delta_0 \eta q \\
\Delta_0 \eta & \varepsilon_{0,0}+\Delta_0 & -i\Delta_0 \eta q & 0 & -i\Delta_0 \eta q & 0 \\
0 & i\Delta_0 \eta q & \varepsilon_{0,1}+\Delta_0 & \Delta_0 \eta (1-q^2) & 0 & -\Delta_0 \eta q^2 \\
-i\Delta_0 \eta q & 0 & \Delta_0 \eta (1-q^2) & \varepsilon_{0,1}+\Delta_0 & -\Delta_0 \eta q^2 & 0 \\
0 & i\Delta_0 \eta q & 0 & -\Delta_0 \eta q^2 & \varepsilon_{0,1}+\Delta_0 & \Delta_0 \eta (1-q^2) \\
-i\Delta_0 \eta q & 0 & -\Delta_0 \eta q^2 & 0 & \Delta_0 \eta (1-q^2)  & \varepsilon_{0,1}+\Delta_0 
\end{pmatrix} , \label{eq:sixsix}
\end{equation}
where we have used previous definitions, including Eq.~(\ref{eq:etapar}).

Diagonalizing $H$ gives the effective energy spectrum, and the resulting ground state gap,
\begin{equation}
E_g\simeq 2\Delta_0\eta q^2+\frac{1}{2} 
\left[ \sqrt{[\hbar\omega_0+2\Delta_0\eta(1-q^2)]^2+2(2\Delta_0\eta q)^2}
- \sqrt{[\hbar\omega_0-2\Delta_0\eta(1-q^2)]^2+2(2\Delta_0\eta q)^2} \right] .
\end{equation}
We can also compute the mixing fraction:
\begin{equation}
{\cal F}_{\tilde{n}=0}\simeq \frac{(2\Delta_0\eta q)^2}
{\left[ \hbar\omega_0+2\Delta_0\eta(1-q^2) \right]^2
+2(2\Delta_0\eta q)^2
+ \left[ \hbar\omega_0+2\Delta_0\eta(1-q^2) \right]
\sqrt{[\hbar\omega_0+2\Delta_0\eta(1-q^2)]^2+2(2\Delta_0\eta q)^2} } .
\end{equation}
\end{widetext}
These results can be compared to an elliptical dot in the high aspect ratio limit (Sec.~\ref{sec:long}).
Making the substitutions $\omega_y\rightarrow \omega_0$ and $\varphi\rightarrow \pi/2$ in Eqs.~(\ref{eq:Evaspect}) and (\ref{eq:Faspect}), consistent with our choice of $\omega_x\gg\omega_y$, we obtain identical results for $E_g$ and ${\cal F}_{\tilde{n}=0}$, up to leading order.  Details of the energy spectrum begin to differ only for higher energy levels. 

\section{Perpendicular Field Geometry} \label{sec:perpendicular}
\subsection{Circular dot} \label{sec:perpcirc}
In contrast with the previous section, the magnetic field plays a non-trivial role for valley coupling in the perpendicular field geometry.
Here, we consider a single electron in a circularly-symmetric parabolic confinement potential,
in a perpendicular magnetic field, $\mathbf{B}=B \hat{\mathbf{z}}$.  (We continue to use unprimed coordinate notation.)  The unperturbed Hamiltonian is given by
\begin{eqnarray}  & & \hspace{-.2in}
H_0=\sum_{j=x,y,z} \frac{1}{2m_j} \left[ -i\hbar \frac{\partial}{\partial r_j} 
+e A_j(\mathbf{r}) \right]^2 
\label{eq:schrorot}
\\ & & \hspace{0.9in} 
+ V_\text{QW}(z) + \frac{1}{2}m_t\omega_0^2(x^2+y^2)  , \nonumber
\end{eqnarray}
where $\omega_0$ characterizes the 
parabolic potential.  We adopt the symmetric gauge for the
vector potential:  $\mathbf{A}(\mathbf{r})=(-By,Bx,0)$.
In this case, the eigenfunction function is separable:
\begin{equation}
F(\mathbf{r})=f(x,y)\zeta(z) . \label{eq:separ}
\end{equation}
Appropriate solutions for the quantum well envelope function, $\zeta(z)$, are discussed elsewhere.\cite{daviesbook,harrisonbook}  As before, we adopt the lowest subband approximation.  
We also choose the position of the interface, $z_0=0$, to simplify our calculations.

The lateral eigenstates $f(x,y)$ are known as Fock-Darwin states.\cite{fock,darwin} 
The properly normalized Fock-Darwin functions can be expressed in terms of 
radial coordinates, $(r\cos \theta,r\sin \theta)=(x,y)$, as follows:
\begin{eqnarray}  \label{eq:FD}
f_{nl}(r,\theta ) &=& \sqrt{\frac{n!b}{2\pi l_B^2 2^{|l|}(n+|l|)!}}
\\ & & \times
e^{il\theta} e^{-br^2/4l_B^2}
\left( \frac{b r^2}{l_B^2} \right)^{|l|/2} \!
L_n^{|l|} \left( \frac{b r^2}{2l_B^2} \right) . \nonumber
\end{eqnarray}
Here, $n$ and $l$ are the orbital and angular momentum quantum numbers, respectively.  $n$ and $l$ are both integers, with $n\geq 0$.  Additionally, $l_B=\sqrt{\hbar/|eB|}$ is the quantum magnetic length scale, and $L_n^l(x)$ is a generalized Laguerre polynomial.\cite{stegun}  The quantity
$b=\sqrt{1+(2\omega_0/\omega_c)^2}$ is defined in terms of the cyclotron and confinement frequencies.  As a consequence of the circular symmetry, we can choose the step direction to lie along $\hat{\bf x}$ without loss of
generality, so that $\varphi =0$.  The Fock-Darwin energy levels are then given by 
\begin{eqnarray}
&& \hspace{-.2in} 
\varepsilon_{nl}=\frac{\hbar \omega_z}{2}  +\frac{\hbar \omega_c}{2} \left[ b(2n+|l|+1)+l \right] 
\label{eq:epperp} \\ \nonumber && \hspace{1.2in} \text{(perpendicular \ $B$-field)} ,
\end{eqnarray}
where we have included the subband energy for completeness, although it will be ignored in the following calculations.

Consistent with previous sections, the intra-valley perturbation matrix elements are given by
\begin{equation}
\langle n',l',v| H | n,l,v \rangle = \delta_{n,n'}\delta_{l,l'}(\varepsilon_{nl}+ \Delta_0) ,
\label{eq:intrav}
\end{equation}
while the inter-valley matrix elements are given by
\begin{widetext}
\begin{eqnarray}
\langle n',l',\bar{v}| V_v | n,l,v \rangle 
&=& v_v\zeta_0^2\frac{b}{2\pi l_B^2}
\sqrt{\frac{n!(n')!}{2^{|l|+|l'|}(n+|l|)!(n'+|l'|)!}}
\label{eq:bigME}  \\ && \hspace{0.25in} \nonumber \times
\int_0^{2\pi}\! d\theta \int_0^{\infty} \! rdr\, 
e^{i(l-l')\theta -2ik_0v\vartheta r\cos \theta
-br^2/2l_B^2} 
\left( \frac{b r^2}{l_B^2} \right)^{(|l|+|l'|)/2} \!\!
L_{n}^{|l|}\left( \frac{br^2}{2l_B^2} \right)
L_{n'}^{|l'|}\left( \frac{br^2}{2l_B^2} \right) 
\\ \nonumber
&=& \left(-ik_0vl_B\vartheta \sqrt{2/b} \right)^{|l-l'|}
v_v\zeta_0^2
\sqrt{n!(n')!(n+|l|)!(n'+|l'|)!} \, e^{-2(k_0l_B\vartheta)^2/b}
\\ && \hspace{0.25in} \nonumber \times
\sum_{\kappa=0}^n \sum_{\kappa'=0}^{n'}
\frac{p!(-1)^{\kappa +\kappa'}L_p^{|l-l'|}(2k_0^2l_B^2\vartheta^2/b)}
{(n-\kappa)!(n'-\kappa')!(|l|+\kappa)!(|l'|+\kappa')!\kappa ! \kappa' !} .
\end{eqnarray}
\end{widetext}
Here, we have defined $2p=|l|+|l'|-|l-l'|+2\kappa +2\kappa'$, and made use of the series expansion for the Laguerre polynomial.\cite{stegun}

\begin{figure}[t] 
  \centering
  \includegraphics[width=2.in,keepaspectratio]{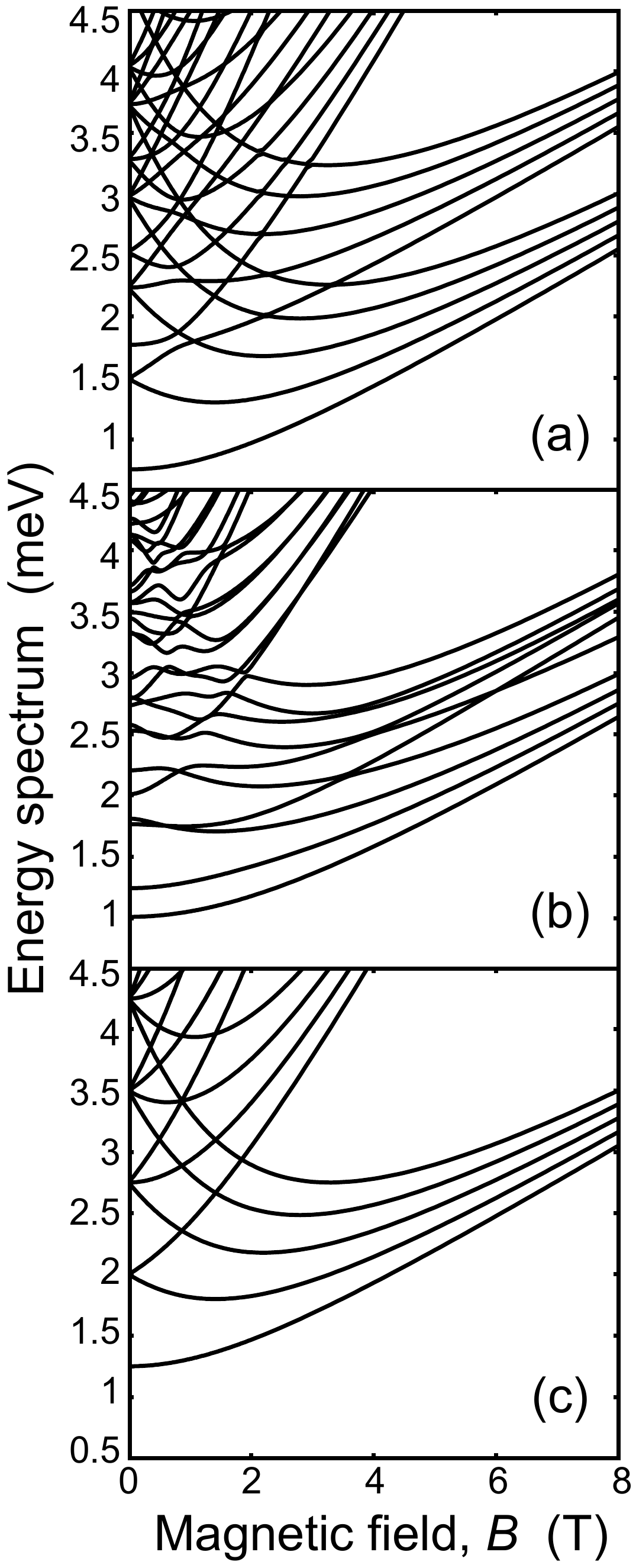}
  \caption{
Energy levels for a circular quantum dot in a perpendicular field geometry, including valley coupling.  Dot parameters are appropriate for a single-electron spin qubit:  $\hbar \omega_0=0.75$~meV, $v_v\zeta_0^2=0.5$~meV.  Uniform step orientation is set to $\varphi =\pi/4$.  The effective miscut tilt angles are given by (a) $\vartheta=0.03^\circ$, (b) $\vartheta=0.3^\circ$,  and (c) $\vartheta=3^\circ$.  (b) is probably closest to experimental conditions.  In (c), each curve is doubly degenerate.}
  \label{fig:Bfield}
\end{figure}

The Hamiltonian matrix elements (\ref{eq:intrav}) and (\ref{eq:bigME}) can be solved numerically.  We obtain a quantum dot energy spectrum for typical experimental conditions, as shown in Fig.~\ref{fig:Bfield}.  The device parameters are similar to those used in Fig.~\ref{fig:elltilt2}.  Note that we have limited the analysis to orbitals in the range $(2n+|l|)<5$.  
The different types of behavior observed in Fig.~\ref{fig:Bfield} are easily understood.  In (c), the interface tilt is large, causing a strong suppression of the valley coupling.  Indeed, every energy level is doubly-degenerate.  The apparent energy spectrum is equivalent to a circular dot in a single-valley material.  In (b), a smaller, but physically realistic, interfacial tilt lifts the orbital and valley degeneracies, and causes a broadening of the high-field energy ``bands."  In this case, the tilt angle is still large enough that the valley splitting is smaller than the orbital splitting at zero field.  We observe a complex assortment of level crossings and anti-crossings.  In (a), the interfacial tilt is small enough that the valley splitting is larger than the orbital splitting at zero field.  At high fields, the levels separate into two distinct subbands.

The emergence of well-defined valley states is not ubiquitous in Fig.~\ref{fig:Bfield}.  By ``well-defined" here, we mean that several states have identical valley characteristics, so that the valley index remains a good quantum number, even after the perturbation.  Well-defined valley states are uncommon because valley-orbit coupling mixes the unperturbed orbitals differently for each eigenstate.  However, when inter-orbital valley coupling is weak, well-defined valley characteristics can emerge.  

Weak inter-orbital valley coupling typically occurs in two ways.  First, it can occur when the valley splitting is inherently small compared to the orbital splitting, for example when the interface tilt is large.  In this case, the intra-orbital valley coupling (\textit{i.e.}, between states with the same orbital quantum numbers) is small, and the inter-orbital valley coupling is much smaller.  The valley potential then produces weak pair-wise couplings between states with the same orbital quantum numbers, as in Fig.~\ref{fig:Bfield}(c).  In this case, the valley splitting is different for each pair of states.  For example, in the $n=0$ manifold (the lowest ``band" in the high-field regime), we find that
\begin{equation}
E_g\simeq 2\Delta_0 \eta L_{|l|}(q^2) , \label{eq:Egperpcirc}
\end{equation}
which depends on the quantum number $l$.  Here, $\Delta_0$ is the same as before, but
\begin{eqnarray}
q &=& k_0\vartheta l_B\sqrt{2/b} , \label{eq:q2} \\
\eta &=& e^{-q^2} , \label{eq:eta2}
\end{eqnarray}
and $L_n(x)\equiv L_n^0(x)$ is an ordinary Laguerre polynomial.  The valley eigenstates have an identical character for all orbital states, however, and can be classified as either ``even" or ``odd."  Similar behavior is observed for large interfacial tilt angles in Fig.~\ref{fig:elltilt2}.  When valley and orbital splittings are comparable in size, as in Figs.~\ref{fig:Bfield}(a) and (b), the inter-orbital coupling competes with the intra-orbital coupling, leading to a more complicated valley-orbit mixing.  

Well-defined valley states also emerge in the high-field limit, as observed in Fig.~\ref{fig:Bfield}(a).  Within a given band, we find that valley coupling occurs primarily between states with the same angular momentum quantum number, and that the valley splitting is identical for all such pairs.  For states with different $l$, Eq.~(\ref{eq:bigME}) shows that the valley coupling is proportional to $(l_B/\sqrt{b})^{|l-l'|}$, which vanishes at high fields.  Thus, in the $n=0$ manifold, we observe pairs of states with the energy splitting $2\Delta_0\eta$, with many intervening states between them.  In other words, two subbands emerge, with a uniform splitting.  The ground state gap between the two lowest levels is then determined by orbital effects, rather than the valley coupling.  Within a given subband, the eigenstates all have an identical valley character:  one subband is ``even," the other is ``odd."

In spite of the rather complicated behavior observed in Fig.~\ref{fig:Bfield}(a)-(c), a simple expression can be given for the splitting between the two lowest energy levels.  We first note that the two lowest states in Fig.~\ref{fig:Bfield}(c) are nearly degenerate, with orbital quantum numbers $n=l=0$.  In Fig.~\ref{fig:Bfield}(b), the low-lying states both have an $n=l=0$ character at low fields.  However, the higher state crosses over to an $n=0$, $l=-1$ character at high fields.  In Fig.~\ref{fig:Bfield}(a), the first excited state has an $n=0$, $l=-1$ character over the entire field range.  From such considerations, we see that the ground state gap should be well characterized within the manifold spanned by quantum numbers $n=0$, $l=\{-1,0\}$, and $v=\pm 1$.  Evaluating the perturbation Hamiltonian in this manifold again leads to Eq.~(\ref{eq:fourfour}), where we now use the definitions (\ref{eq:epperp}), (\ref{eq:q2}), and (\ref{eq:eta2}).  The results for the ground state gap, Eq.~(\ref{eq:Evaspect}), and the mixing fraction, Eq.~(\ref{eq:Faspect}), then apply, if we make the replacement
\begin{equation}
\hbar \omega_y \rightarrow (\hbar \omega_c/2)(b-1) .
\end{equation}
We emphasize that these results correctly interpolate between orbital and valley dominated behavior, over the entire field range.

\subsection{Elliptical Dot}
To close this section, we investigate an elliptical quantum dot in a perpendicular magnetic field.  We cannot directly apply the results of Sec.~\ref{sec:parallel}.  However, the problem is still
separable with respect to the $z'$ coordinate, as in Eq.~(\ref{eq:separ}).  The magnetic field now couples the two transverse directions, $\hat{\bf x}'$ and $\hat{\bf y}'$, in a fundamental way.  The transverse effective mass is nearly isotropic
when the interfacial tilt angle $\vartheta$ is small:\cite{friesen07}  $m_x\simeq m_y\simeq m_t$.
So it is unnecessary to rescale the length axes.  Dropping the prime notation, the transverse envelope function is given by
\begin{widetext}
\begin{eqnarray}
f_{n_x,n_y}(x,y) &=& N_{n_x,n_y} \exp \left\{ \frac{iJ xy}{\hbar} 
-\frac{M_1\omega_1x^2+2iL M_1\omega_1M_2\omega_2xy+M_2\omega_2y^2}
{2\hbar (L^2M_1\omega_1M_2\omega_2+1)} \right\}
\\ \nonumber && \hspace{-.5in} \times
\sum_{k=0}^{n_x}\sum_{l=0}^{n_y} \frac{c_{k,l}}{k!l!(n_x-k)!(n_y-l)!} H_{n_x-k} \left[
\frac{\sqrt{2M_1\omega_1/\hbar}(x+iL M_2\omega_2y)}
{L^2M_1\omega_1M_2\omega_2+1} \right]
H_{n_y-l} \left[
\frac{\sqrt{2M_2\omega_2/\hbar}(L M_1\omega_1x-iy)}
{L^2M_1\omega_1M_2\omega_2+1} \right] ,
\end{eqnarray}
where, now,
\end{widetext}
\begin{eqnarray}
J &=& -\frac{m_t}{2} \left( \frac{\omega_x^2-\omega_y^2}{\omega_c} \right)
\\ && \nonumber
+\frac{m_t}{2\omega_c}
\sqrt{(\omega_x^2+\omega_y^2+\omega_c^2)^2-4\omega_x^2\omega_y^2} , 
\end{eqnarray}
\begin{equation}
L = \frac{\omega_c}{m_t 
\sqrt{(\omega_x^2+\omega_y^2+\omega_c^2)^2-4\omega_x^2\omega_y^2}} , 
\end{equation}
\begin{eqnarray}
M_{1,2} &=& 
\\ \nonumber && \hspace{-.3in}
\frac{2m_t \sqrt{(\omega_x^2+\omega_y^2+\omega_c^2)^2-4\omega_x^2\omega_y^2}}
{(\omega_x^2-\omega_y^2\pm \omega_c^2)+
\sqrt{(\omega_x^2+\omega_y^2+\omega_c^2)^2-4\omega_x^2\omega_y^2}} , 
\end{eqnarray}
\begin{eqnarray}
\omega_{1,2} &=& \frac{1}{\sqrt{2}}
\bigglb[ \omega_x^2+\omega_y^2+\omega_c^2
\\ \nonumber && \hspace{.3in}
\pm \sqrt{(\omega_x^2+\omega_y^2+\omega_c^2)^2-4\omega_x^2\omega_y^2} \biggrb]^{1/2} , 
\end{eqnarray}
\begin{equation}
N_{n_x,n_y} = \frac{(-i)^{n_y}(3/4)^{1/4}}{2^{n_x+n_y}} \left[
\frac{(n_x)!(n_y)! \sqrt{M_1\omega_1M_2\omega_2}}{\hbar(L^2M_1\omega_1M_2\omega_2+1) }
\right]^{1/2} \hspace{-.1in} , 
\end{equation}
\begin{widetext}
\begin{equation}
c_{k,l} = \left\{ 
\begin{array}{lc}
(-1)^{-l/2}12^{(l+k)/4}
(3-4D^2G^2)^{(l-k-1)/4}(1-2D^2)^{(k-l)/2} 
P_{(k-l-1)/2}^{(k+l+1)/2} \left(
\sqrt{\frac{4D^2G^2}{4D^2G^2-3}} \right) 
& \text{($k+l$ even),} \\
0 & \text{($k+l$ odd),} \end{array} 
\right. 
\end{equation}
\end{widetext}
\begin{equation}
D = \frac{2L^2M_1\omega_1M_2\omega_2}{L^2M_1\omega_1M_2\omega_2+1} ,
\end{equation}
\begin{equation}
G = -\frac{2}{L^2M_1\omega_1M_2\omega_2+1} .
\end{equation}
Note that $n_x$ and $n_y$ are non-negative integers, $P_\nu^\mu (z)$ is an associated Legendre function, \cite{stegun} and we have assumed $\omega_x\geq\omega_y$.
 
The energy eigenvalues are given by
\begin{equation}
\varepsilon_{n_x,n_y}=\frac{\hbar \omega_z}{2}+\left(n_x+\frac{1}{2} \right)\hbar\omega_1
+\left(n_y+\frac{1}{2} \right)\hbar\omega_2 . \label{eq:epbig}
\end{equation}
As before, we have adopted the lowest subband approximation.

We can compute the valley coupling matrix elements.  The intra-valley terms are given by
\begin{equation}
\langle n_x'n_y'v|H|n_xn_yv\rangle
=\delta_{n_x,n_x'}\delta_{n_y,n_y'}
[\varepsilon_{n_x,n_y}+v_v\zeta_0^2] ,
\end{equation}
as in Eq.~(\ref{eq:MEdiag0}).
The inter-valley terms are given by
\begin{widetext}
\begin{eqnarray}
&& \langle n_x'n_y'\bar{v}|H|n_xn_yv\rangle
=v_v\zeta_0^2e^{2ik_0z_0v} \frac{\pi}{2} \left( \frac{v}{2} \right)^{n_x+n_x'+n_y+n_y'}
\sqrt{3(n_x)!(n_x')!(n_y)!(n_y')!}
\label{eq:bigell} \\ \nonumber && \hspace{.5in} \times
\exp \left[ -\hbar k_0^2\vartheta^2(Q+1)
\left( \frac{\cos^2\varphi}{M_1\omega_1} +\frac{\sin^2\varphi}{M_2\omega_2} \right) \right]
\sum_{k=0}^{n_x} \;\; \sum_{l=0}^{n_y} \;\; \sum_{k'=0}^{n_x'} \;\, \sum_{l'=0}^{n_y'} \;
\sum_{m=0}^{n_x-k} \sum_{q=0}^{n_y-l} 
\sum_{m'=0}^{n_x'-k'} \sum_{q'=0}^{n_y'-l'} 
\sum_{p=0}^{\lfloor m/2\rfloor} \sum_{r=0}^{\lfloor q/2\rfloor} 
\sum_{p'=0}^{\lfloor m'/2\rfloor} \sum_{r'=0}^{\lfloor q'/2\rfloor} 
 \\ \nonumber && \hspace{.5in} \times 
 \Biggl\{ c_{k,l}c_{k',l'} (i)^{-n_y+n_y'+q-q'-m+m'}(-1)^{n_x+n_y'+k+l'}
 Q^{u/2}\left(\frac{2}{Q+1}\right)^{t/2}
  \\ \nonumber && \hspace{0.6in} \frac{ \hspace{1.45in} \times 
 H_s\left(k_0\vartheta\cos \varphi\sqrt{\hbar(Q+1)/M_1\omega_1}\right)
  H_{t-s}\left(k_0\vartheta\sin \varphi\sqrt{\hbar(Q+1)/M_2\omega_2}\right) \Biggr\} }
  { \hspace{-1.5in} \Bigl[ (k)!(k')!(l)!(l')!(p)!(p')!(r)!(r')! (m-2p)!(m'-2p')!(q-2r)!(q'-2r')! } ,
   \\ \nonumber && \hspace{2.05in} \times 
  (n_x-k-m)!(n_x'-k'-m')!(n_y-l-q)!(n_y'-l'-q')! \Bigr] 
\end{eqnarray} 
\end{widetext}
where we have defined
\begin{eqnarray}
s &=& m+m'+q+q'-2(p+p'+r+r') ,\\
t &=& n_x+n_x'+n_y+n_y'-k-k'-l-l' \nonumber \\
&& -2(p+p'+r+r') ,\\
u &=& n_x+n_x'-k-k'-m-m'+q+q' \nonumber \\
&& -2(r+r'),
\end{eqnarray}
and
\begin{equation}
Q=\frac{\omega_c^2M_1\omega_1M_2\omega_2/m_t^2}
{(\omega_x^2+\omega_y^2+\omega_c^2)^2-4\omega_x^2\omega_y^2} ,
\end{equation}
and we have made use of the series expansion for Hermite polynomials.\cite{stegun}

The valley coupling for an elliptical quantum dot in a perpendicular field geometry, Eq.~(\ref{eq:bigell}), is the most general solution obtained in this paper.  The resulting energy spectra are qualitatively similar to the geometries studied above, and we do not plot them here.  

We consider two limiting cases.  In the small dot limit, the ground state can be treated using the same perturbation Hamiltonian as Eq.~(\ref{eq:twotwo}).  However, $\varepsilon_0$ now corresponds to $\varepsilon_{0,0}$ in Eq.~(\ref{eq:epbig}), and
\begin{equation}
\eta = \exp \left[ -\hbar k_0^2\vartheta^2 (Q+1)
\left( \frac{ \cos^2 \varphi}{M_1\omega_1}+\frac{ \sin^2 \varphi}{M_2\omega_2} \right) \right]  , 
\label{eq:d1} 
\end{equation} 
while $\Delta_0$ is unchanged.  The ground state gap is then given by
\begin{equation}
E_g \simeq  2\Delta_0\eta \label{eq:Evellperplim} .
\end{equation}
In the further limit of $\omega_x=\omega_y$, we recover the equivalent result for the circular quantum dot in a perpendicular field geometry:
\begin{equation}
E_g=2v_v\zeta_0^2 \exp [-2(k_0l_B\vartheta )^2/b] .
\end{equation}
Alternatively, keeping $\omega_x\neq \omega_y$, but taking the limit $B\rightarrow 0$, we recover Eq.~(\ref{eq:orbital}), corresponding to a small elliptical quantum dot in the parallel field geometry.

We also consider the important case where the valley coupling is of the same order as the lowest orbital splitting ($\Delta_0\eta \sim \hbar \omega_y \lesssim \hbar\omega_x$).  Analogous to Sec.~\ref{sec:long}, we compute the Hamiltonian in the manifold spanned by the ordered basis $\{ |0 0 +\rangle,|0 0 -\rangle,|0 1 +\rangle,|0 1 -\rangle\}$, obtaining
\begin{widetext}
\begin{equation}
H\simeq \begin{pmatrix}
\varepsilon_{0,0}+\Delta_0 & \Delta_0 \eta & 0 & \Delta_0 \eta(-r+iq) \\
\Delta_0 \eta & \varepsilon_{0,0}+\Delta_0 & \Delta_0 \eta(r-iq) & 0 \\
0 & \Delta_0 \eta(r+iq) & \varepsilon_{0,1}+\Delta_0 & \Delta_0 \eta (1-r^2-q^2) \\
\Delta_0 \eta(-r-iq) & 0 & \Delta_0 \eta (1-r^2-q^2) & \varepsilon_{0,1}+\Delta_0
\end{pmatrix} , \label{eq:fourfourperp}
\end{equation}
\end{widetext}
where the various parameters are defined above, except
\begin{eqnarray}
q &=& (k_0\vartheta \sin \varphi ) \sqrt{2\hbar /M_2\omega_2 } , \label{eq:d2} \\
r &=& (k_0\vartheta \cos \varphi ) \sqrt{2\hbar Q/M_1\omega_1 } . \label{eq:d3}
\end{eqnarray}
We can diagonalize the Hamiltonian and compute the ground state gap and the mixing fraction as before.  This yields identical results to Eqs.~(\ref{eq:Evaspect}) and (\ref{eq:Faspect}), if we replace $q^2\rightarrow (q^2+r^2)$, and use the parameter definitions given above.  In the further limit $B\rightarrow 0$, we find that $r=0$, and the parameters defined in this section reduce to those of Sec.~\ref{sec:long}.  The results are then equivalent.

\section{Discussion and Conclusions} \label{sec:discussion}
The term ``valley splitting" is widely used in the literature of indirect gap semiconductors.  However, its meaning is not precise.  As demonstrated above, valley-orbit coupling can lead to a complete mixing of orbital wavefunctions in the regime of interest for quantum computing, so that pure valley and orbital quantum number are no longer good quantum numbers.  When atomic steps are present at the quantum well interface, many different ``valley splittings" can be observed in the energy spectrum, none of which are universal.  

We propose some formal definitions to help clarify this situation.  First, we define the ``characteristic" or theoretical maximum of the valley splitting, corresponding to $2\Delta_0$.  This splitting is achieved in a hypothetical flat quantum well, and its value is independent of the quantum dot shape.  Second, we define a ``renormalized" valley splitting, corresponding to $2\Delta_0\eta$.  This quantity accounts for the suppression caused by a tilted interface.  The renormalization factor $\eta$ depends on the dot shape, but it does not include the effects of inter-orbital valley coupling.  Finally, we note that the energy splitting between the lowest two quantum dot eigenstates should not, accurately, be referred to as the valley splitting, since this usage is only valid in limiting cases.  We have identified some of those cases here (\textit{e.g.}, the small dot limit and the high-field limit).  More accurately, we should simply refer to the ``ground state gap," which is of particular interest for quantum computing and other applications.  In this paper, we have obtained analytical expressions for the ground state gap and for the mixing fraction, in several cases of interest.

Our main conclusions for the valley coupling in quantum dots can be summarized as follows:  the shape and the size of the quantum dot, the number of steps that it covers at the quantum well interface, and the nature of the disorder all affect the suppression of the valley splitting.  The trends in the energy spectrum of Fig.~\ref{fig:elltilt2} are difficult to characterize in a simple way.  One might expect excited orbitals to exhibit a larger suppression because of their larger size.  However, the effect is obscured by the nodal structure of the wavefunctions.  The behavior is more plain when we compare ground state wavefunctions of different sizes, as in Fig.~\ref{fig:radius}.  In this case, smaller dots clearly exhibit a larger valley coupling.  As expected, the effect depends sensitively on the tilt angle of the interface.  The dependence of the valley coupling on dot size is also manifested in Fig.~\ref{fig:Bfield}(a), at high fields.  Here, the interface tilt angle is rather small, and the magnetic confinement of the  quantum dot is strong, so tilting has little effect on the energy spectrum.  The valley splitting then approaches its theoretical maximum, regardless of the specific dot shape.

Experiments in Si 2DEGs generally corroborate our results.  Examples include Hall bar measurements, where the valley splitting grows monotonically with the magnetic field.\cite{goswami07, khrapai03,wilde05}  This behavior is consistent with the case of small or vanishing electrostatic confinement in a quantum dot [\textit{e.g.}, the case $l=0$, $b\simeq 1$ in Eq.~(\ref{eq:Egperpcirc})].  In a quantum point contact, valley splitting is significantly enhanced due to lateral confinement.\cite{goswami07}  In a Si/SiGe two-electron quantum dot, a large singlet-triplet splitting between 0.1 and 0.3~meV has been measured at zero-field.\cite{shaji08}  Since the triplet state involves an excited single-electron orbital, the singlet-triplet splitting provides a good estimate of the ground state gap.  This solitary data point is therefore consistent with Fig.~\ref{fig:Bfield}(b).  Due to valley-orbit coupling, it is not appropriate to identify the splitting as purely orbital-like or valley-like.  However, small gaps of any type are anathema for qubit operations.  The large value of the measured gap implies an acceptable splitting for qubit applications.\cite{shaji08}  In future experiments, it would be desirable to perform excited state spectroscopy in similar dots, as a function of magnetic field.  The present theory would then allow the valley coupling and other quantum dot confinement parameters to be determined.

In this paper, we have paid special attention to the inter-orbital valley coupling.   The origin of this coupling can be understood in terms of the dot shape.  According to the variational principle, a quantum dot will minimize its ground state energy by maximizing its valley splitting.  In our perturbation theory, the ground state will therefore mix with asymmetric excited states that allow it to squeeze along the direction of the step gradient.  The mixing occurs only between opposing valley states, $v$ and $\bar{v}$, whose coupling is the source of valley splitting.  

The inter-orbital component of valley coupling has not been emphasized previously.  In quantum information applications, it leads to a potential source of decoherence for spin qubits.  When phonons are present, valley scattering can occur in a process analogous to spin flip transitions.\cite{tahan02}  Two separate ingredients are needed to flip a spin by a phonon:\cite{hasagawa,roth}  spin-orbit coupling (to mix spin states via the excited orbitals) and a magnetic field (to lift the Kramers degeneracy).  Valley scattering can occur in the presence of two analogous processes:  valley-orbit coupling at a tilted interface (to mix valley states via the excited orbitals) and a sharp confinement potential (to lift the valley degeneracy).

In the context of quantum computing, valley scattering does not directly affect the spin state or lead to spin decoherence.  However, there is an indirect spin effect, mediated by the Pauli principle, which depends on the valley coupling.   The predominant two-spin gate interaction is the exchange coupling,\cite{loss98} which occurs because any two electrons must have different quantum numbers. When the valley degree of freedom is introduced, the size of the Hilbert space doubles, providing an opportunity for an electron to ``leak" into a non-qubit sector via valley scattering.  If this happens, the electron will have a different quantum number, and the exchange coupling will be ineffective.  The spin state is then uncontrollable.  Aside from this mechanism, a direct spin-valley coupling has also been predicted for Si 2DEGs, with a magnitude comparable to the spin-orbit coupling.\cite{nestoklon06,nestoklon08}  We do not consider this effect here.

Finally, we note that the results obtained in this paper were obtained by assuming a smooth, uniformly tilted interface.  We expect such results to pertain to more general situations.  However, step disorder and discreteness can both have a quantitative effect on the valley splitting.  In the Appendix, we discuss several experimental conditions that are inconsistent with our approximations, including:  (i) step discreteness, which becomes important for widely separated steps; inhomogeneous tilting, which affects (ii) the quantum dot confinement potential and (iii) the delicate phenomenon of destructive interference in the valley coupling; (iv) step bunching, which also tends to moderate the destructive interference.  Since all of these phenomena enhance the valley splitting, we can treat them, phenomenologically, through an effective tilt angle $\vartheta$ that is smaller than the nominal tilt.

In the absence of any detailed knowledge of the interfacial disorder, $\vartheta$ must remain a phenomenological parameter.  However, the relation between the average tilt and the effective tilt, and the dependence on different forms of disorder, remains an important topic for future investigation.  For example, the energy levels of a Si double quantum dot could potentially be measured as a function of magnetic field by the method of Ref.~\onlinecite{shaji08}.  Comparison with the present theory can then provide information about $\vartheta$.  More intriguingly perhaps, it may be possible to use gating methods to shift the center of a dot, and thereby map out $\vartheta$ as a function of position.

In conclusion, we have developed a theory of valley coupling in realistic quantum dots.  The resulting energy spectra exhibit crossings and anti-crossings, as a function of the interfacial tilt angle and magnetic field.  Due to valley-orbit coupling, the ground state gap is not strictly orbital-like or valley-like, except in certain limiting cases.  For quantum dots of interest in quantum computing, we find that inter-valley orbital coupling plays an important role in device operation, and in determining the ground state gap.

\begin{acknowledgments}
We are indebted to M.A. Eriksson and R. Joynt for helpful discussions about this work and for comments on the manuscript.  The work was supported by NSA/ARO Contract No.\ W911NF-08-1-0482, NSF Grant Nos.\ DMR-0325634, CCF-0523675, CCF-0523680, and DMR-0805045, and Sandia National Laboratory Agreement No.\ 649787.  Sandia is a multiprogram laboratory operated by Sandia Corporation, a Lockheed Martin Company, for the US DOE under Contract No.\ DE-AC04-94AL85000.  This research is part of Sandia's Laboratory Directed Research and Development (LDRD) program.
M.F. would also like to acknowledge support from the University of New South Wales, where some of this work was completed.
\end{acknowledgments}

\appendix
\section{Validity of Approximations Used in the Calculations}
This appendix addresses the approximations made in the main text that enable analytic solutions to be obtained, and it discusses the conditions under which they apply.    The main approximations are that:  (1) the interfacial tilt can be treated as continuous and the presence of individual atomic steps can be ignored, and (2) the disorder in the step locations can be ignored.
The first approximation suggests that there will be no relaxation of the electron wavefunction to conform to the pattern of steps at the interface.  The second approximation is equivalent to assuming the interfacial tilt is smooth and uniform.

\subsection{Step Discreteness: Phase Relaxation at Discrete Steps}
In this work, we treat the interface as continuous and we ignore the presence of individual atomic steps.  Naively, one would expect such an approximation to be reasonable because the step is much smaller than the characteristic size of the envelope function.  For example, the height of a single atomic step along [001] is $a/4=1.4$~\AA, while a quantum dot may extend over 10's of nm.  

Energy arguments suggest that such an argument breaks down when the interface tilt angle is small.  In Eq.~(\ref{eq:Kohn}), we have taken the complex phase difference between $\alpha_{\tilde{n},n,v}$ and $\alpha_{\tilde{n},n,\bar{v}}$ to be a constant, independent of the lateral position.  The global value of the phase difference is determined by minimizing the ground state energy for the whole quantum dot, according to degenerate perturbation theory.  Since the optimal value of the phase difference depends on the interface position $z_0$, it is clear that valley coupling cannot be optimized simultaneously on both sides of a step while the phase remains constant.  By allowing the phase to vary, we can reduce this valley coupling energy penalty, but we pay a price in kinetic energy.  Careful energy balance calculations show that phase relaxation becomes favorable when $\vartheta \leq 0.1-0.3^\circ$.  In the present work, phase relaxation would tend to reduce valley-orbit mixing.  

\subsection{Disorder-Induced Confinement Potentials}
Step disorder causes non-uniform tilting of the interface, and leads to spatial variations of the valley splitting and the quantum dot confinement potential.  The position of the electron will shift when the local curvature of the valley splitting is larger than the curvature of the electrostatic potential.  The net effect of the shift is to reduce the effective tilt angle.

\subsection{Disorder-Induced Symmetry Reduction}
For highly uniform step distributions, the suppression of the valley coupling by valley phase interference is very effective.  However, this interference effect is somewhat delicate.  The reduction of symmetry by step disorder can lead to an order of magnitude increase in the valley splitting.\cite{friesen06}  Related effects are also observed in the presence of alloy disorder in the SiGe quantum well barriers.\cite{kharche07}  While step disorder tends to reduce the effective tilt angle,\cite{friesen06} the net effect of alloy disorder appears to be more complicated.\cite{kharche07,srinivasan08}

\subsection{Step Bunching}
Interfacial steps may become bunched due to randomness, or as a consequence of strained growth.\cite{tersoff95}  The suppression of valley splitting, as discussed in this paper, is particularly effective for single-atom steps, as compared to bunched steps.  We can understand this as follows.  From Eq.~(\ref{eq:Kohn}) and the related discussion, the phase difference between the $z$ valleys for a flat interface
is  $2k_0z_0$.  At a step of height $a/4$, we see that $z_0\rightarrow z_0+a/4$, so the phase difference suddenly changes by $0.82\pi$.  Since the two sides of the step are almost fully out of phase, there is a significant suppression of the total valley splitting.  On the other hand, a two-fold bunched step of height $a/2$ has a much milder effect on the valley coupling.  In fact, the first step geometry with a stronger effect on the valley coupling is a six-fold bunched step.  In this sense, single-atom steps are essentially a worst-case scenario, leading to maximal suppression of the valley splitting.  The dependence of the effective tilt angle on step bunching is therefore rather complicated.

\subsection{Summary}
To summarize, the assumption that the steps can be treated as continuous is expected to be valid, but valley phase relaxation and variability in the step density both serve to modify the effective tilt angle of the interface.  Assuming that
an appropriate effective tilt angle is used, we expect our analytical expressions to provide an accurate depiction of the valley coupling for typical quantum dot geometries.  Our results therefore provide a useful first approximation for understanding the valley coupling.  Several numerical results confirm this point of view.\cite{friesen06,kharche07,srinivasan08}

To obtain a quantitative estimate for the valley coupling, one must know the effective interfacial tilt angle $\vartheta$.  For the reasons described above, the effective tilt depends in a nontrivial way on the average tilt angle, the disorder in the step configuration, and the materials properties of the interface.  A quantitative calculation of $\vartheta$ is particularly challenging, since the dependence of the valley coupling on the tilt angle is exponential, so that errors and fluctuations in the tilt angle are essentially magnified.  For example, in Eq.~(\ref{eq:varthetas}), an exponential suppression of the valley coupling was predicted for tilt angles larger than $0.26^\circ$, based on realistic dot parameters.  Experimentally, a large valley splitting has been observed for wavefunctions in a quantum point contact with a $2^\circ$ miscut\cite{goswami07}  and in quantum dots in samples with nominal tilt angles approaching $2^\circ$(Ref.~\onlinecite{shaji08}) --
evidence that the effective tilt in these devices is significantly smaller than the nominal tilt.


\begin{thebibliography}{99}
\bibitem{noteosc}
Examples include the matrix element $\langle n',\bar{v}|H_0|n,v\rangle$, which is off-diagonal in the valley index.  Such terms are small because the integrand involves a product of slowly varying envelope functions with fast sine or cosine oscillations.
\end{thebibliography}
\end{document}